# Operation of a Modular 3D-Pixelated Liquid Argon Time-Projection Chamber in a Neutrino Beam

S. Abbaslu [1], A. Abed Abud [2], R. Acciarri [3], L. P. Accorsi [4], M. A. Acero [5], M. R. Adames [4], G. Adamov [6], M. Adamowski [3], C. Adriano [7], F. Akbar [8], F. Alemanno [9], N. S. Alex [8], K. Allison [10], M. Alrashed [11], A. Alton [12], R. Alvarez [13], T. Alves [14], A. Aman [15], H. Amar [16], P. Amedo [16,17], J. Anderson [18], D. A. Andrade [19], C. Andreopoulos [20], M. Andreotti [21,22], M. P. Andrews [3], F. Andrianala [23], S. Andringa [24], F. Anjarazafy [23], S. Ansarifard [1], D. Antic [25], M. Antoniassi [4], A. Aranda-Fernandez [26], L. Arellano [27], E. Arrieta Diaz [28], M. A. Arroyave [3], M. Arteropons [29], J. Asaadi [30], M. Ascencio [31], A. Ashkenazi [32], D. Asner [33], L. Asquith [34], E. Atkin [14], D. Auguste [35], A. Aurisano [36], V. Aushev [37], D. Autiero [38], D. Ávila Gómez [39], M. B. Azam [19], F. Azfar [40], A. Back [41], J. J. Back [42], Y. Bae [43], I. Bagaturia [6], L. Bagby [3], D. Baigarashev [44], S. Balasubramanian [3], A. Balboni [21,22], P. Baldi [45], W. Baldini [21], J. Baldonedo [46], B. Baller [3], B. Bambah [47], F. Barao [24,48], D. Barbu [49], G. Barenboim [16], P. Barham Alzás [2], G. J. Barker [42], W. Barkhouse [50], G. Barr [40], A. Barros [4], N. Barros [24,51], D. Barrow [40], J. L. Barrow [43], A. Basharina-Freshville [52], A. Bashyal [33], V. Basque [3], M. Bassani [53], D. Basu [54], C. Batchelor [55], L. Bathe-Peters [40], J. B. R. Battat [56], F. Battisti [57], J. Bautista [43], F. Bay [58], J. L. L. Bazo Alba [59], J. F. Beacom [60], E. Bechetoille [38], B. Behera [61], E. Belchior [62], B. Bell [63], G. Bell [64], L. Bellantoni [3], G. Bellettini [65,66], V. Bellini [67,68], O. Beltramello [2], A. Belyaev [69], C. Benitez Montiel [16,70], D. Benjamin [33], F. Bento Neves [24], J. Berger [71], S. Berkman [72], J. Bermudez [73], J. Bernal [70], P. Bernardini [9,74], A. Bersani [75], E. Bertholet [32], E. Bertolini [76], S. Bertolucci [57,77], M. Betancourt [3], A. Betancur Rodríguez [39], Y. Bezawada [78], A. T. Bezerra [79], A. Bhat [80], V. Bhatnagar [81], M. Bhattacharjee [82], S. Bhattacharjee [62], M. Bhattacharya [3], S. Bhuller [40], B. Bhuyan [82], S. Biagi [83], J. Bian [45], K. Biery [3], B. Bilki [84,85], M. Bishai [33], A. Blake [86], F. D. Blaszczyk [3], G. C. Blazey [54], E. Blucher [80], B. Bogart [87], J. Boissevain [88], S. Bolognesi [89], T. Bolton [11], L. Bomben [76,90], M. Bonesini [76,91], C. Bonilla-Diaz [92], A. Booth [93], F. Boran [41], R. Borges Merlo [7], N. Bostan [85], G. Botogoske [94], B. Bottino [75,95], R. Bouet [96], J. Boza [71], J. Bracinik [97], B. Brahma [98], D. Brailsford [86], F. Bramati [76], A. Branca [76], A. Brandt [30], J. Bremer [2], S. J. Brice [3], V. Brio [67], C. Brizzolari [76,91], C. Bromberg [72], J. Brooke [25], A. Bross [3], G. Brunetti [76,91], M. B. Brunetti [99], N. Buchanan [71], H. Budd [8], J. Buergi [100], A. Bundock [25], D. Burgardt [101], S. Butchart [34], G. Caceres V. [78], R. Calabrese [94], R. Calabrese [21,22], J. Calcutt [33,102], L. Calivers [100], E. Calvo [13], A. Caminata [75], A. F. Camino [103], W. Campanelli [24], A. Campani [75,95], A. Campos Benitez [104], N. Canci [94], J. Capó [16], I. Caracas [105], D. Caratelli [106], D. Carber [71], J. M. Carceller [2], G. Carini [33], B. Carlus [38], M. F. Carneiro [33], P. Carniti [76,91], I. Caro Terrazas [71], H. Carranza [30], N. Carrara [78], L. Carroll [11], T. Carroll [107], A. Carter [108], E. Casarejos [46], D. Casazza [21], J. F. Castaño Forero [109], F. A. Castaño [110], C. Castromonte [111], E. Catano-Mur [112], C. Cattadori [76], F. Cavalier [35], F. Cavanna [3], S. Centro [29], G. Cerati [3], C. Cerna [113], A. Cervelli [57], A. Cervera Villanueva [16], J. Chakrani [114], M. Chalifour [2], A. Chappell [42], A. Chatterjee [115], B. Chauhan [85], C. Chavez Barajas [20], H. Chen [33], M. Chen [45], W. C. Chen [116], Y. Chen [117], Z. Chen [45], D. Cherdack [118], S. S. Chhibra [93], C. Chi [119], F. Chiapponi [57], R. Chirco [19], N. Chitirasreemadam [65,66], K. Cho [120], S. Choate [85], G. Choi [8], D. Chokheli [6], P. S. Chong [121], B. Chowdhury [18], D. Christian [3], M. Chung [122], E. Church [123], M. F. Cicala [52], M. Cicerchia [29], V. Cicero [57,77], R. Ciolini [65], P. Clarke [55], G. Cline [114], A. G. Cocco [94], J. A. B. Coelho [124], A. Cohen [124], J. Collazo [46], J. Collot [125],

Check for updates

H. Combs [104], J. M. Conrad [126], L. Conti [127], T. Contreras [3], M. Convery [117], K. Conway [128], S. Copello [129], P. Cova [53,130], C. Cox [108], L. Cremonesi [93], J. I. Crespo-Anadón [13], M. Crisler [3], E. Cristaldo [70,76], J. Crnkovic [3], G. Crone [52], R. Cross [42], A. Cudd [10], C. Cuesta [13], Y. Cui [131], F. Curciarello [132], D. Cussans [25], J. Dai [125], O. Dalager [3], W. Dallaway [116], R. D′Amico [21,22], H. da Motta [133], Z. A. Dar [112], R. Darby [34], L. Da Silva Peres [134], Q. David [38], G. S. Davies [135], S. Davini [75], J. Dawson [124], R. De Aguiar [7], P. Debbins [85], M. P. Decowski [136,137], A. de Gouvêa [138], P. C. De Holanda [7], P. De Jong [136,137], P. Del Amo Sanchez [139], G. De Lauretis [38], A. Delbart [89], M. Delgado [76,91], A. Dell′Acqua [2], G. Delle Monache [132], N. Delmonte [53,130], P. De Lurgio [18], R. Demario [72], G. De Matteis [9,74], J. R. T. de Mello Neto [134], A. P. A. De Mendonca [7], D. M. DeMuth [140], S. Dennis [141], C. Densham [142], P. Denton [33], G. W. Deptuch [33], A. De Roeck [2], V. De Romeri [16], J. P. Detje [141], J. Devine [2], K. Dhanmeher [38], R. Dharmapalan [143], M. Dias [144], A. Diaz [145], J. S. Díaz [41], F. Díaz [59], F. Di Capua [94,146], A. Di Domenico [147,148], S. Di Domizio [75,95], S. Di Falco [65], L. Di Giulio [2], P. Ding [3], L. Di Noto [75,95], E. Diociaiuti [132], G. Di Sciascio [127], V. Di Silvestre [147], C. Distefano [83], R. Di Stefano [127], R. Diurba [100], M. Diwan [33], Z. Djurcic [18], S. Dolan [2], M. Dolce [101], M. J. Dolinski [63], D. Domenici [132], S. Dominguez [13], S. Donati [65,66], S. Doran [31], D. Douglas [117], T. A. Doyle [128], F. Drielsma [117], D. Duchesneau [139], K. Duffy [40], K. Dugas [45], P. Dunne [14], B. Dutta [149], D. A. Dwyer [114], A. S. Dyshkant [54], S. Dytman [103], M. Eads [54], A. Earle [34], S. Edayath [31], D. Edmunds [72], J. Eisch [3], W. Emark [54], P. Englezos [150], A. Ereditato [80], T. Erjavec [78], C. O. Escobar [3], J. J. Evans [27], E. Ewart [41], A. C. Ezeribe [151], K. Fahey [3], A. Falcone [76,91], M. Fani′ [43,88], D. Faragher [43], C. Farnese [73], Y. Farzan [1], J. Felix [152], Y. Feng [31], M. Ferreira da Silva [144], G. Ferry [35], E. Fialova [153], L. Fields [154], P. Filip [155], A. Filkins [156], F. Filthaut [136,157], G. Fiorillo [94,146], M. Fiorini [21,22], S. Fogarty [71], W. Foreman [88], J. Fowler [158], J. Franc [153], K. Francis [54], D. Franco [80], J. Franklin [159], J. Freeman [3], J. Fried [33], A. Friedland [117], M. Fucci [128], S. Fuess [3], I. K. Furic [160], K. Furman [93], A. P. Furmanski [43], R. Gaba [81], A. Gabrielli [57,77], A. M Gago [59], F. Galizzi [76,91], H. Gallagher [161], M. Galli [124], N. Gallice [33], V. Galymov [38], E. Gamberini [2], T. Gamble [151], R. Gandhi [162], S. Ganguly [3], F. Gao [106], S. Gao [33], D. Garcia-Gamez [163], M. Á. García-Peris [27], S. Gardiner [3], A. Gartman [153], A. Gauch [100], P. Gauzzi [147,148], S. Gazzana [132], G. Ge [119], N. Geffroy [139], B. Gelli [7], S. Gent [164], L. Gerlach [33], A. Ghosh [31], T. Giammaria [21,22], D. Gibin [29,73], I. Gil-Botella [13], A. Gioiosa [127], S. Giovannella [132], A. K. Giri [98], V. Giusti [65], D. Gnani [114], O. Gogota [37], S. Gollapinni [88], K. Gollwitzer [3], R. A. Gomes [165], L. S. Gomez Fajardo [166], D. Gonzalez-Diaz [17], J. Gonzalez-Santome [2], M. C. Goodman [18], S. Goswami [115], C. Gotti [76], J. Goudeau [62], C. Grace [114], E. Gramellini [27], R. Gran [167], P. Granger [2], C. Grant [168], D. R. Gratieri [7,169], G. Grauso [94], P. Green [40], S. Greenberg [114,170], W. C. Griffith [34], K. Grzelak [171], L. Gu [86], W. Gu [33], V. Guarino [18], M. Guarise [21,22], R. Guenette [27], M. Guerzoni [57], D. Guffanti [76,91], A. Guglielmi [73], F. Y. Guo [128], A. Gupta [172], V. Gupta [136,137], G. Gurung [30], D. Gutierrez [173], P. Guzowski [27], M. M. Guzzo [7], S. Gwon [174], A. Habig [167], L. Haegel [38], R. Hafeji [16,17], L. Hagaman [80], A. Hahn [3], J. Hakenmüller [158], T. Hamernik [3], P. Hamilton [14], J. Hancock [97], M. Handley [141], F. Happacher [132], B. Harris [121], D. A. Harris [3,175], L. Harris [143], A. L. Hart [93], J. Hartnell [34], T. Hartnett [142], J. Harton [71], T. Hasegawa [176], C. M. Hasnip [2], R. Hatcher [3], S. Hawkins [72], J. Hays [93], M. He [118], A. Heavey [3], K. M. Heeger [177], A. Heindel [128], J. Heise [178], P. Hellmuth [96], L. Henderson [102], K. Herner [3], V. Hewes [36], A. Higuera [179], A. Himmel [3], E. Hinkle [80], L. R. Hirsch [4], J. Ho [180], J. Hoefken Zink [57], J. Hoff [3], A. Holin [142], T. Holvey [40], C. Hong [124], S. Horiuchi [104], G. A. Horton-Smith [11], R. Hosokawa [181], T. Houdy [35], B. Howard [3,175], R. Howell [8], I. Hristova [142], M. S. Hronek [3], H. Hua [14], J. Huang [78], R. G. Huang [114], X. Huang [135], Z. Hulcher [117], A. Hussain [11], G. Iles [14], N. Ilic [116], A. M. Iliescu [132], R. Illingworth [3], G. Ingratta [175], A. Ioannisian [69],

M. Ismerio Oliveira [134], C. M. Jackson [123], V. Jain [182], E. James [3], W. Jang [30], B. Jargowsky [45], D. Jena [3], I. Jentz [107], C. Jiang [183], J. Jiang [128], A. Jipa [49], J. H. Jo [33], F. R. Joaquim [24,48], W. Johnson [61], C. Jollet [96], R. Jones [151], N. Jovancevic [184], M. Judah [103], C. K. Jung [128], K. Y. Jung [8], T. Junk [3], Y. Jwa [117,119], M. Kabirnezhad [14], A. C. Kaboth [108,142], I. Kadenko [37], O. Kalikulov [44], D. Kalra [119], M. Kandemir [185], S. Kar [25], G. Karagiorgi [119], G. Karaman [85], A. Karcher [114], Y. Karyotakis [139], S. P. Kasetti [62], L. Kashur [71], A. Kauther [54], N. Kazaryan [69], L. Ke [33], E. Kearns [168], P. T. Keener [121], K. J. Kelly [149], R. Keloth [104], E. Kemp [7], O. Kemularia [6], Y. Kermaidic [35], W. Ketchum [3], S. H. Kettell [33], N. Khan [14], A. Khvedelidze [6], D. Kim [149], J. Kim [8], M. J. Kim [3], S. Kim [174], B. King [3], M. King [80], M. Kirby [33], A. Kish [3], J. Klein [121], J. Kleykamp [135], A. Klustova [14], T. Kobilarcik [3], L. Koch [105], K. Koehler [107], L. W. Koerner [118], D. H. Koh [117], M. Kordosky [112], T. Kosc [125], V. A. Kostelecký [41], I. Kotler [63], W. Krah [136], R. Kralik [34], M. Kramer [114], F. Krennrich [31], T. Kroupova [121], S. Kubota [27], M. Kubu [2], V. A. Kudryavtsev [151], G. Kufatty [15], S. Kuhlmann [18], A. Kumar [43], J. Kumar [143], M. Kumar [172], P. Kumar [186], P. Kumar [151], S. Kumaran [45], J. Kunzmann [100], V. Kus [153], T. Kutter [62], J. Kvasnicka [155], T. Labree [54], M. Lachat [8], T. Lackey [3], I. Lalău [49], A. Lambert [114], B. J. Land [121], C. E. Lane [63], N. Lane [27], K. Lang [187], T. Langford [177], M. Langstaff [27], F. Lanni [2], J. Larkin [8], P. Lasorak [14], D. Last [8], A. Laundrie [107], G. Laurenti [57], E. Lavaut [35], H. Lay [86], I. Lazanu [49], R. LaZur [71], M. Lazzaroni [53,188], S. Leardini [17], J. Learned [143], T. LeCompte [117], G. Lehmann Miotto [2], R. Lehnert [41], M. Leitner [114], H. Lemoine [167], D. Leon Silverio [61], L. M. Lepin [15], J.-Y. Li [55], S. W. Li [45], Y. Li [33], R. Lima [79], C. S. Lin [114], D. Lindebaum [25], S. Linden [33], R. A. Lineros [92], A. Lister [107], B. R. Littlejohn [19], J. Liu [45], Y. Liu [80], S. Lockwitz [3], I. Lomidze [6], K. Long [14], J. Lopez [110], I. López de Rego [13], N. López-March [16], J. M. LoSecco [154], A. Lozano Sanchez [63], X.-G. Lu [42], K. B. Luk [114,170,189], X. Luo [106], E. Luppi [21,22], A. A. Machado [7], P. Machado [3], C. T. Macias [41], J. R. Macier [3], M. MacMahon [52], S. Magill [18], C. Magueur [35], K. Mahn [72], A. Maio [24,51], N. Majeed [11], A. Major [158], K. Majumdar [20], A. Malige [119], S. Mameli [65], M. Man [116], R. C. Mandujano [45], J. Maneira [24,51], S. Manly [8], K. Manolopoulos [142], M. Manrique Plata [41], S. Manthey Corchado [13], L. Manzanillas-Velez [139], E. Mao [156], M. Marchan [3], A. Marchionni [3], D. Marfatia [143], C. Mariani [104], J. Maricic [143], F. Marinho [190], A. D. Marino [10], T. Markiewicz [117], F. Das Chagas Marques [7], M. Marshak [43], C. M. Marshall [8], J. Marshall [42], L. Martina [9,74], J. Martín-Albo [16], D. A. Martinez Caicedo [61], M. Martinez-Casales [3], F. Martínez López [41], S. Martynenko [33], V. Mascagna [76], A. Mastbaum [150], M. Masud [174], F. Matichard [114], G. Matteucci [94,146], J. Matthews [62], C. Mauger [121], N. Mauri [57,77], K. Mavrokoridis [20], I. Mawby [86], F. Mayhew [72], T. McAskill [56], N. McConkey [93], B. McConnell [41], K. S. McFarland [8], C. McGivern [3], C. McGrew [128], A. McNab [27], C. McNulty [114], J. Mead [136], L. Meazza [76], V. C. N. Meddage [160], A. Medhi [82], M. Mehmood [175], B. Mehta [81], P. Mehta [186], F. Mei [57,77], P. Melas [191], L. Mellet [72], T. C. D. Melo [79], O. Mena [16], H. Mendez [173], D. P. Méndez [33], A. Menegolli [129,192], G. Meng [73], A. C. E. A. Mercuri [4], A. Meregaglia [96], M. D. Messier [41], S. Metallo [43], W. Metcalf [62], M. Mewes [41], H. Meyer [101], T. Miao [3], J. Micallef [126,161], A. Miccoli [9], G. Michna [164], R. Milincic [143], F. Miller [107], G. Miller [27], W. Miller [43], A. Minotti [76,91], L. Miralles Verge [2], C. Mironov [124], S. Miscetti [132], C. S. Mishra [3], P. Mishra [47], S. R. Mishra [193], D. Mladenov [2], I. Mocioiu [194], A. Mogan [3], R. Mohanta [47], T. A. Mohayai [41], N. Mokhov [3], J. Molina [70], L. Molina Bueno [16], E. Montagna [57,77], A. Montanari [57], C. Montanari [3,129,192], D. Montanari [3], D. Montanino [9,74], L. M. Montaño Zetina [195], M. Mooney [71], A. F. Moor [151], M. Moore [117], Z. Moore [156], D. Moreno [109], G. Moreno-Granados [104], O. Moreno-Palacios [112], L. Morescalchi [65], C. Morris [118], E. Motuk [52], C. A. Moura [196], G. Mouster [86], W. Mu [3], L. Mualem [145], J. Mueller [3], M. Muether [101], A. Muir [64], Y. Mukhamejanov [44], A. Mukhamejanova [44], M. Mulhearn [78],

D. Munford [118], L. J. Munteanu [2], H. Muramatsu [43], J. Muraz [125], M. Murphy [104], T. Murphy [156], A. Mytilinaki [142], J. Nachtman [85], Y. Nagai [197], S. Nagu [198], D. Naples [103], S. Narita [181], J. Nava [57,77], A. Navrer-Agasson [14,27], N. Nayak [33], M. Nebot-Guinot [55], A. Nehm [105], J. K. Nelson [112], O. Neogi [85], J. Nesbit [107], M. Nessi [2,3], D. Newbold [142], M. Newcomer [121], D. Newmark [126], R. Nichol [52], F. Nicolas-Arnaldos [163], A. Nielsen [45], A. Nikolica [121], J. Nikolov [184], E. Niner [3], X. Ning [33], K. Nishimura [143], A. Norman [3], A. Norrick [3], P. Novella [16], A. Nowak [86], J. A. Nowak [86], M. Oberling [18], J. P. Ochoa-Ricoux [45], S. Oh [158], S. B. Oh [3], A. Olivier [154], T. Olson [118], Y. Onel [85], Y. Onishchuk [37], A. Oranday [41], M. Osbiston [42], J. A. Osorio Vélez [110], L. O'Sullivan [105], L. Otiniano Ormachea [111,199], L. Pagani [78], G. Palacio [39], O. Palamara [3], S. Palestini [200], J. M. Paley [3], M. Pallavicini [75,95], C. Palomares [13], S. Pan [115], M. Panareo [9,74], P. Panda [47], V. Pandey [3], W. Panduro Vazquez [108], E. Pantic [78], V. Paolone [103], A. Papadopoulou [88], R. Papaleo [83], D. Papoulias [191], S. Paramesvaran [25], J. Park [174], S. Parke [3], S. Parsa [100], S. Parveen [186], M. Parvu [49], D. Pasciuto [65], S. Pascoli [57,77], L. Pasqualini [57,77], J. Pasternak [14], G. Patel [43], J. L. Paton [3], C. Patrick [55], L. Patrizii [57], R. B. Patterson [145], T. Patzak [124], A. Paudel [3], J. Paul [136], L. Paulucci [190], Z. Pavlovic [3], G. Pawloski [43], D. Payne [20], A. Peake [108], V. Pec [155], E. Pedreschi [65], S. J. M. Peeters [34], W. Pellico [3], E. Pennacchio [38], A. Penzo [85], O. L. G. Peres [7], Y. F. Perez Gonzalez [159], L. Pérez-Molina [13], C. Pernas [112], J. Perry [55], D. Pershey [15], G. Pessina [76], G. Petrillo [117], C. Petta [67,68], R. Petti [193], M. Pfaff [14], V. Pia [57,77], G. M. Piacentino [127], L. Pickering [108,142], L. Pierini [21,22], F. Pietropaolo [2,73], V. L. Pimentel [7,201], G. Pinaroli [33], S. Pincha [82], J. Pinchault [139], K. Pitts [104], P. Plesniak [14], K. Pletcher [72], K. Plows [40], C. Pollack [173], T. Pollmann [136,137], F. Pompa [16], X. Pons [2], N. Poonthottathil [31,172], V. Popov [32], F. Poppi [57,77], J. Porter [34], L. G. Porto Paixão [7], M. Potekhin [33], M. Pozzato [57,77], R. Pradhan [98], T. Prakash [114], M. Prest [76], F. Psihas [3], D. Pugnere [38], D. Pullia [2,124], X. Qian [33], J. Queen [158], J. L. Raaf [3], M. Rabelhofer [41], V. Radeka [33], J. Rademacker [25], F. Raffaelli [65], A. Rafique [18], A. Rahe [54], S. Rajagopalan [33], M. Rajaoalisoa [36], I. Rakhno [3], L. Rakotondravohitra [23], M. A. Ralaikoto [23], L. Ralte [98], M. A. Ramirez Delgado [121], B. Ramson [3], S. S. Randriamanampisoa [23], A. Rappoldi [129,192], G. Raselli [129,192], T. Rath [61], P. Ratoff [86], R. Ray [3], H. Razafinime [36], R. F. Razakamiandra [128], E. M. Rea [43], J. S. Real [125], B. Rebel [3,107], R. Rechenmacher [3], J. Reichenbacher [61], S. D. Reitzner [3], E. Renner [88], S. Repetto [75,95], S. Rescia [33], F. Resnati [2], C. Reynolds [93], M. Ribas [4], S. Riboldi [53], C. Riccio [128], G. Riccobene [83], J. S. Ricol [125], M. Rigan [34], A. Rikalo [184], E. V. Rincón [39], A. Ritchie-Yates [108], D. Rivera [88], A. Robert [125], A. Roberts [20], E. Robles [45], M. Roda [20], D. Rodas Rodríguez [17], M. J. O. Rodrigues [79], J. Rodriguez Rondon [61], S. Rosauro-Alcaraz [35], P. Rosier [35], D. Ross [72], M. Rossella [129,192], M. Ross-Lonergan [119], T. Rotsy [23], N. Roy [175], P. Roy [101], P. Roy [104], C. Rubbia [202], D. Rudik [94], A. Ruggeri [57], G. Ruiz Ferreira [27], K. Rushiya [186], B. Russell [126], S. Sacerdoti [124], N. Saduyev [44], S. K. Sahoo [98], N. Sahu [98], S. Sakhiyev [44], P. Sala [3], G. Salmoria [4], S. Samanta [75], M. C. Sanchez [15], A. Sánchez-Castillo [163], P. Sanchez-Lucas [163], D. A. Sanders [135], S. Sanfilippo [83], D. Santoro [53,130], N. Saoulidou [191], P. Sapienza [83], I. Sarcevic [203], I. Sarra [132], G. Savage [3], V. Savinov [103], G. Scanavini [177], A. Scanu [76], A. Scaramelli [129], T. Schefke [62], H. Schellman [3,102], S. Schifano [21,22], P. Schlabach [3], D. Schmitz [80], A. W. Schneider [126], K. Scholberg [158], A. Schroeder [43], A. Schukraft [3], B. Schuld [10], S. Schwartz [145], A. Segade [46], E. Segreto [7], A. Selyunin [100], C. R. Senise [144], J. Sensenig [121], S. H. Seo [3], D. Seppela [72], M. H. Shaevitz [119], P. Shanahan [3], P. Sharma [81], R. Kumar [204], S. Sharma Poudel [61], K. Shaw [34], T. Shaw [3], K. Shchablo [38], J. Shen [121], C. Shepherd-Themistocleous [142], J. Shi [141], W. Shi [128], S. Shin [205], S. Shivakoti [101], A. Shmakov [45], I. Shoemaker [104], D. Shooltz [72], R. Shrock [128], M. Siden [71], J. Silber [114], L. Simard [35], J. Sinclair [117], G. Sinev [61], Jaydip Singh [78], J. Singh [198], L. Singh [206],

**P. Singh [93], V. Singh [206], S. Singh Chauhan [81], R. Sipos [2], C. Sironneau [124], G. Sirri [57], K. Siyeon [174], K. Skarpaas [117], J. Smedley [8], J. Smith [128], P. Smith [41], J. Smolik [153,155], M. Smy [45], M. Snape [42], E. L. Snider [3], P. Snopok [19], M. Soares Nunes [3], H. Sobel [45], M. Soderberg [156], H. Sogarwal [31], C. J. Solano Salinas [207], S. Söldner-Rembold [14], N. Solomey [101], V. Solovov [24], W. E. Sondheim [88], M. Sorbara [127], M. Sorel [16], J. Soto-Oton [16], A. Sousa [36], K. Soustruznik [208], D. Souza Correia [133], F. Spinella [65], J. Spitz [87], N. J. C. Spooner [151], D. Stalder [70], M. Stancari [3], L. Stanco [29,73], J. Steenis [78], R. Stein [25], H. M. Steiner [114], A. F. Steklain Lisbôa [4], J. Stewart [33], B. Stillwell [80], J. Stock [61], T. Stokes [177], T. Strauss [3], L. Strigari [149], A. Stuart [26], J. G. Suarez [39], J. Subash [97], A. Surdo [9], L. Suter [3], A. Sutton [15], K. Sutton [145], Y. Suvorov [94,146], R. Svoboda [78], S. K. Swain [209], C. Sweeney [31], B. Szczerbinska [210], A. M. Szelc [55], A. Sztuc [52], A. Taffara [65], N. Talukdar [193], J. Tamara [109], H. A. Tanaka [117], S. Tang [33], N. Taniuchi [141], A. M. Tapia Casanova [211], A. Tapper [14], S. Tariq [3], E. Tatar [212], R. Tayloe [41], A. M. Teklu [128], K. Tellez Giron Flores [33], J. Tena Vidal [32], P. Tennessen [58,114], M. Tenti [57], K. Terao [117], F. Terranova [76,91], G. Testera [75], T. Thakore [36], A. Thea [142], S. Thomas [156], A. Thompson [138], C. Thorpe [27], S. C. Timm [3], E. Tiras [85,185], V. Tishchenko [33], S. Tiwari [8], N. Todorović [184], L. Tomassetti [21,22], A. Tonazzo [124], D. Torbunov [33], D. Torres Muñoz [61], M. Torti [76,91], M. Tortola [16], Y. Torun [19], N. Tosi [57], D. Totani [71], M. Toups [3], C. Touramanis [20], V. Trabattoni [53], D. Tran [118], J. Trevor [145], E. Triller [72], S. Trilov [25], D. Trotta [76], J. Truchon [107], D. Truncali [147,148], W. H. Trzaska [213], Y. Tsai [45], Y.-T. Tsai [117], Z. Tsamalaidze [6], K. V. Tsang [117], N. Tsverava [6], S. Z. Tu [183], S. Tufanli [2], C. Tunnell [179], J. Turner [159], M. Tuzi [16], M. Tzanov [62], M. A. Uchida [141], J. Ureña González [16], J. Urheim [41], T. Usher [117], H. Utaegbulam [8], S. Uzunyan [54], M. R. Vagins [45,214], P. Vahle [112], G. A. Valdiviesso [79], E. Valencia [152], R. Valentim [144], Z. Vallari [60], E. Vallazza [76], J. W. F. Valle [16], R. Van Berg [121], D. V. Forero [211], A. Vannozzi [132], M. Van Nuland-Troost [136], F. Varanini [73], D. Vargas Oliva [116], N. Vaughan [102], K. Vaziri [3], A. Vázquez-Ramos [163], J. Vega [199], J. Vences [24,51], S. Ventura [73], A. Verdugo [13], M. Verzocchi [3], K. Vetter [3], M. Vicenzi [33], H. Vieira de Souza [124], C. Vignoli [215], C. Vilela [24], E. Villa [2], S. Viola [83], B. Viren [33], G. V. Stenico [55], R. Vizarreta [8], A. P. Vizcaya Hernandez [71], S. Vlachos [27], G. Vorobyev [193], Q. Vuong [8], A. V. Waldron [93], L. Walker [118], H. Wallace [108], M. Wallach [72], J. Walsh [72], T. Walton [3], L. Wan [3], B. Wang [85], H. Wang [216], J. Wang [61], M. H. L. S. Wang [3], X. Wang [3], Y. Wang [217], D. Warner [71], L. Warsame [142], M. O. Wascko [40,142], D. Waters [52], A. Watson [97], K. Wawrowska [34,142], A. Weber [3,105], C. M. Weber [43], M. Weber [100], H. Wei [62], A. Weinstein [31], S. Westerdale [131], M. Wetstein [31], K. Whalen [142], A. J. White [80,*], L. H. Whitehead [141], D. Whittington [156], F. Wieler [4], J. Wilhlemi [177], M. J. Wilking [43], A. Wilkinson [42], C. Wilkinson [114], F. Wilson [142], R. J. Wilson [71], P. Winter [18], J. Wolcott [161], J. Wolfs [8], T. Wongjirad [161], A. Wood [118], K. Wood [114], E. Worcester [33], M. Worcester [33], K. Wresilo [141], M. Wright [27], M. Wrobel [71], S. Wu [43], W. Wu [45], Z. Wu [45], M. Wurm [105], J. Wyenberg [180], B. M. Wynne [55], Y. Xiao [45], I. Xiotidis [14], B. Yaeggy [36], N. Yahlali [16], E. Yandel [106], G. Yang [33,128], J. Yang [189], T. Yang [3], A. Yankelevich [45], L. Yates [3], U. Yevarouskaya [128], K. Yonehara [3], T. Young [50], B. Yu [33], H. Yu [33], J. Yu [30], W. Yuan [55], M. Zabloudil [153], R. Zaki [175], J. Zalesak [155], L. Zambelli [139], B. Zamorano [163], A. Zani [53], O. Zapata [110], L. Zazueta [156], G. P. Zeller [3], J. Zennamo [3], J. Zettlemoyer [3], K. Zeug [107], C. Zhang [33], S. Zhang [41], Y. Zhang [33], L. Zhao [45], M. Zhao [33], E. D. Zimmerman [10], S. Zucchelli [57,77], V. Zutshi [54], R. Zwaska [3] and on behalf of the DUNE Collaboration**

1 Institute for Research in Fundamental Sciences, Tehran, Iran
2 CERN, The European Organization for Nuclear Research, 1211 Meyrin, Switzerland
3 Fermi National Accelerator Laboratory, Batavia, IL 60510, USA
4 Universidade Tecnológica Federal do Paraná, Curitiba, Brazil
5 Universidad del Atlántico, Barranquilla, Atlántico, Colombia

[6] Georgian Technical University, Tbilisi, Georgia
[7] Universidade Estadual de Campinas, Campinas 13083-970, Brazil
[8] University of Rochester, Rochester, NY 14627, USA
[9] Istituto Nazionale di Fisica Nucleare Sezione di Lecce, 73100 Lecce, Italy
[10] University of Colorado Boulder, Boulder, CO 80309, USA
[11] Kansas State University, Manhattan, KS 66506, USA
[12] Augustana University, Sioux Falls, SD 57197, USA
[13] CIEMAT, Centro de Investigaciones Energéticas, Medioambientales y Tecnológicas, E-28040 Madrid, Spain
[14] Imperial College of Science, Technology and Medicine, London SW7 2BZ, UK
[15] Florida State University, Tallahassee, FL 32306, USA
[16] Instituto de Física Corpuscular, CSIC and Universitat de València, 46980 Paterna, Valencia, Spain
[17] Instituto Galego de Física de Altas Enerxías, University of Santiago de Compostela, 15782 Santiago de Compostela, Spain
[18] Argonne National Laboratory, Argonne, IL 60439, USA
[19] Illinois Institute of Technology, Chicago, IL 60616, USA
[20] University of Liverpool, Liverpool L69 7ZE, UK
[21] Istituto Nazionale di Fisica Nucleare Sezione di Ferrara, I-44122 Ferrara, Italy
[22] University of Ferrara, Ferrara, Italy
[23] University of Antananarivo, Antananarivo 101, Madagascar
[24] Laboratório de Instrumentação e Física Experimental de Partículas, 1649-003 Lisboa and 3004-516 Coimbra, Portugal
[25] University of Bristol, Bristol BS8 1TL, UK
[26] Universidad de Colima, Colima, Mexico
[27] University of Manchester, Manchester M13 9PL, UK
[28] Universidad del Magdalena, Santa Marta, Colombia
[29] Universtà degli Studi di Padova, I-35131 Padova, Italy
[30] University of Texas at Arlington, Arlington, TX 76019, USA
[31] Iowa State University, Ames, IA 50011, USA
[32] Tel Aviv University, Tel Aviv-Yafo, Israel
[33] Brookhaven National Laboratory, Upton, NY 11973, USA
[34] University of Sussex, Brighton BN1 9RH, UK
[35] Université Paris-Saclay, CNRS/IN2P3, IJCLab, 91405 Orsay, France
[36] University of Cincinnati, Cincinnati, OH 45221, USA
[37] Taras Shevchenko National University of Kyiv, 01601 Kyiv, Ukraine
[38] Institut de Physique des 2 Infinis de Lyon, 69622 Villeurbanne, France
[39] Universidad EIA, Envigado, Antioquia, Colombia
[40] University of Oxford, Oxford OX1 3RH, UK
[41] Indiana University, Bloomington, IN 47405, USA
[42] University of Warwick, Coventry CV4 7AL, UK
[43] University of Minnesota Twin Cities, Minneapolis, MN 55455, USA
[44] Institute of Nuclear Physics at Almaty, Almaty 050032, Kazakhstan
[45] University of California Irvine, Irvine, CA 92697, USA
[46] University of Vigo, E-36310 Vigo, Spain
[47] University of Hyderabad, Gachibowli, Hyderabad 500 046, India
[48] Instituto Superior Técnico—IST, Universidade de Lisboa, 1049-001 Lisboa, Portugal
[49] University of Bucharest, Bucharest, Romania
[50] University of North Dakota, Grand Forks, ND 58202-8357, USA
[51] Faculdade de Ciências da Universidade de Lisboa—FCUL, 1749-016 Lisboa, Portugal
[52] University College London, London WC1E 6BT, UK
[53] Istituto Nazionale di Fisica Nucleare Sezione di Milano, 20133 Milano, Italy
[54] Northern Illinois University, DeKalb, IL 60115, USA
[55] University of Edinburgh, Edinburgh EH8 9YL, UK
[56] Wellesley College, Wellesley, MA 02481, USA
[57] Istituto Nazionale di Fisica Nucleare Sezione di Bologna, 40127 Bologna, BO, Italy
[58] Antalya Bilim University, 07190 Döşemealtı/Antalya, Turkey
[59] Pontificia Universidad Católica del Perú, Lima, Perú
[60] Ohio State University, Columbus, OH 43210, USA
[61] South Dakota School of Mines and Technology, Rapid City, SD 57701, USA
[62] Louisiana State University, Baton Rouge, LA 70803, USA
[63] Drexel University, Philadelphia, PA 19104, USA
[64] Daresbury Laboratory, Cheshire WA4 4AD, UK
[65] Istituto Nazionale di Fisica Nucleare Laboratori Nazionali di Pisa, Pisa, PI, Italy
[66] Università di Pisa, I-56127 Pisa, Italy
[67] Istituto Nazionale di Fisica Nucleare Sezione di Catania, I-95123 Catania, Italy
[68] Università di Catania, 2, 95131 Catania, Italy

[69] Yerevan Institute for Theoretical Physics and Modeling, Yerevan 0036, Armenia
[70] Universidad Nacional de Asunción, San Lorenzo, Paraguay
[71] Colorado State University, Fort Collins, CO 80523, USA
[72] Michigan State University, East Lansing, MI 48824, USA
[73] Istituto Nazionale di Fisica Nucleare Sezione di Padova, 35131 Padova, Italy
[74] Università del Salento, 73100 Lecce, Italy
[75] Istituto Nazionale di Fisica Nucleare Sezione di Genova, 16146 Genova, GE, Italy
[76] Istituto Nazionale di Fisica Nucleare Sezione di Milano Bicocca, 3, I-20126 Milano, Italy
[77] Università di Bologna, 40127 Bologna, Italy
[78] University of California Davis, Davis, CA 95616, USA
[79] Universidade Federal de Alfenas, Poços de Caldas 37715-400, MG, Brazil
[80] University of Chicago, Chicago, IL 60637, USA
[81] Panjab University, Chandigarh 160014, India
[82] Indian Institute of Technology Guwahati, Guwahati 781 039, India
[83] Istituto Nazionale di Fisica Nucleare Laboratori Nazionali del Sud, 95123 Catania, Italy
[84] Beykent University, Istanbul, Turkey
[85] University of Iowa, Iowa City, IA 52242, USA
[86] Lancaster University, Lancaster LA1 4YB, UK
[87] University of Michigan, Ann Arbor, MI 48109, USA
[88] Los Alamos National Laboratory, Los Alamos, NM 87545, USA
[89] IRFU, CEA, Université Paris-Saclay, F-91191 Gif-sur-Yvette, France
[90] University of Insubria, Via Ravasi, 2, 21100 Varese, VA, Italy
[91] Università di Milano Bicocca, 20126 Milano, Italy
[92] Universidad Católica del Norte, Antofagasta, Chile
[93] Queen Mary University of London, London E1 4NS, UK
[94] Istituto Nazionale di Fisica Nucleare Sezione di Napoli, I-80126 Napoli, Italy
[95] Università degli Studi di Genova, Genova, Italy
[96] Laboratoire de Physique des Deux Infinis Bordeaux, IN2P3, F-33175 Gradignan, Bordeaux, France
[97] University of Birmingham, Birmingham B15 2TT, UK
[98] Indian Institute of Technology Hyderabad, Hyderabad 502285, India
[99] University of Kansas, Lawrence, KS 66045, USA
[100] University of Bern, CH-3012 Bern, Switzerland
[101] Wichita State University, Wichita, KS 67260, USA
[102] Oregon State University, Corvallis, OR 97331, USA
[103] University of Pittsburgh, Pittsburgh, PA 15260, USA
[104] Virginia Tech, Blacksburg, VA 24060, USA
[105] Johannes Gutenberg-Universität Mainz, 55122 Mainz, Germany
[106] University of California Santa Barbara, Santa Barbara, CA 93106, USA
[107] University of Wisconsin Madison, Madison, WI 53706, USA
[108] Royal Holloway College London, London TW20 0EX, UK
[109] Universidad Antonio Nariño, Bogotá, Colombia
[110] University of Antioquia, Medellín, Colombia
[111] Universidad Nacional de Ingeniería, Lima 25, Perú
[112] William and Mary, Williamsburg, VA 23187, USA
[113] Particle Physics and Cosmology International Research Laboratory, Chicago, IL 60637, USA
[114] Lawrence Berkeley National Laboratory, Berkeley, CA 94720, USA
[115] Physical Research Laboratory, Ahmedabad 380 009, India
[116] University of Toronto, Toronto, ON M5S 1A1, Canada
[117] SLAC National Accelerator Laboratory, Menlo Park, CA 94025, USA
[118] University of Houston, Houston, TX 77204, USA
[119] Columbia University, New York, NY 10027, USA
[120] Korea Institute of Science and Technology Information, Daejeon 34141, Republic of Korea
[121] University of Pennsylvania, Philadelphia, PA 19104, USA
[122] Ulsan National Institute of Science and Technology, Ulsan 689-798, Republic of Korea
[123] Pacific Northwest National Laboratory, Richland, WA 99352, USA
[124] Université Paris Cité, CNRS, Astroparticule et Cosmologie, Paris, France
[125] University Grenoble Alpes, CNRS, Grenoble INP, LPSC-IN2P3, 38000 Grenoble, France
[126] Massachusetts Institute of Technology, Cambridge, MA 02139, USA
[127] Istituto Nazionale di Fisica Nucleare Roma Tor Vergata, 00133 Roma, RM, Italy
[128] Stony Brook University, SUNY, Stony Brook, NY 11794, USA
[129] Istituto Nazionale di Fisica Nucleare Sezione di Pavia, I-27100 Pavia, Italy
[130] University of Parma, 43121 Parma, PR, Italy
[131] University of California Riverside, Riverside, CA 92521, USA
[132] Istituto Nazionale di Fisica Nucleare Laboratori Nazionali di Frascati, Frascati, Roma, Italy
[133] Centro Brasileiro de Pesquisas Físicas, Rio de Janeiro 22290-180, RJ, Brazil

[134] Universidade Federal do Rio de Janeiro, Rio de Janeiro 21941-901, RJ, Brazil
[135] University of Mississippi, University, MS 38677, USA
[136] Nikhef National Institute of Subatomic Physics, 1098 XG Amsterdam, The Netherlands
[137] University of Amsterdam, NL-1098 XG Amsterdam, The Netherlands
[138] Northwestern University, Evanston, IL 60208, USA
[139] Laboratoire d'Annecy de Physique des Particules, Université Savoie Mont Blanc, CNRS, LAPP-IN2P3, 74000 Annecy, France
[140] Valley City State University, Valley City, ND 58072, USA
[141] University of Cambridge, Cambridge CB3 0HE, UK
[142] STFC Rutherford Appleton Laboratory, Didcot OX11 0QX, UK
[143] University of Hawaii, Honolulu, HI 96822, USA
[144] Universidade Federal de São Paulo, São Paulo 09913-030, Brazil
[145] California Institute of Technology, Pasadena, CA 91125, USA
[146] Università degli Studi di Napoli Federico II, 80138 Napoli, NA, Italy
[147] Sapienza University of Rome, 00185 Roma, RM, Italy
[148] Istituto Nazionale di Fisica Nucleare Sezione di Roma, 00185 Roma, RM, Italy
[149] Texas A&M University, College Station, TX 77840, USA
[150] Rutgers University, Piscataway, NJ 08854, USA
[151] University of Sheffield, Sheffield S3 7RH, UK
[152] Universidad de Guanajuato, Guanajuato C.P. 37000, Mexico
[153] Czech Technical University, 115 19 Prague 1, Czech Republic
[154] University of Notre Dame, Notre Dame, IN 46556, USA
[155] Institute of Physics, Czech Academy of Sciences, 182 00 Prague 8, Czech Republic
[156] Syracuse University, Syracuse, NY 13244, USA
[157] Radboud University, NL-6525 AJ Nijmegen, The Netherlands
[158] Duke University, Durham, NC 27708, USA
[159] Durham University, Durham DH1 3LE, UK
[160] University of Florida, Gainesville, FL 32611-8440, USA
[161] Tufts University, Medford, MA 02155, USA
[162] Harish-Chandra Research Institute, Jhunsi, Allahabad 211 019, India
[163] University of Granada & CAFPE, 18002 Granada, Spain
[164] South Dakota State University, Brookings, SD 57007, USA
[165] Universidade Federal de Goias, Goiania, GO 74690-900, Brazil
[166] Universidad Sergio Arboleda, 11022 Bogotá, Colombia
[167] University of Minnesota Duluth, Duluth, MN 55812, USA
[168] Boston University, Boston, MA 02215, USA
[169] Fluminense Federal University, 9 Icaraí Niterói 24220-900, RJ, Brazil
[170] University of California Berkeley, Berkeley, CA 94720, USA
[171] University of Warsaw, 02-093 Warsaw, Poland
[172] Indian Institute of Technology Kanpur, Uttar Pradesh 208016, India
[173] University of Puerto Rico, Mayaguez, PR 00681, USA
[174] Chung-Ang University, Seoul 06974, Republic of Korea
[175] York University, Toronto, ON M3J 1P3, Canada
[176] High Energy Accelerator Research Organization (KEK), Ibaraki 305-0801, Japan
[177] Yale University, New Haven, CT 06520, USA
[178] Sanford Underground Research Facility, Lead, SD 57754, USA
[179] Rice University, Houston, TX 77005, USA
[180] Dordt University, Sioux Center, IA 51250, USA
[181] Iwate University, Morioka, Iwate 020-8551, Japan
[182] University of Albany, SUNY, Albany, NY 12222, USA
[183] Jackson State University, Jackson, MS 39217, USA
[184] University of Novi Sad, 21102 Novi Sad, Serbia
[185] Erciyes University, Kayseri, Turkey
[186] Jawaharlal Nehru University, New Delhi 110067, India
[187] University of Texas at Austin, Austin, TX 78712, USA
[188] Università degli Studi di Milano, I-20133 Milano, Italy
[189] Hong Kong University of Science and Technology, Kowloon, Hong Kong, China
[190] Instituto Tecnológico de Aeronáutica, Sao Jose dos Campos, Brazil
[191] University of Athens, 157 84 Zografou, GR, Greece
[192] Università degli Studi di Pavia, 27100 Pavia, PV, Italy
[193] University of South Carolina, Columbia, SC 29208, USA
[194] Pennsylvania State University, University Park, PA 16802, USA
[195] Centro de Investigación y de Estudios Avanzados del Instituto Politécnico Nacional (Cinvestav), Mexico City, Mexico
[196] Universidade Federal do ABC, Santo André 09210-580, SP, Brazil

[197] Eötvös Loránd University, 1053 Budapest, Hungary
[198] University of Lucknow, Uttar Pradesh 226007, India
[199] Comisión Nacional de Investigación y Desarrollo Aeroespacial, Lima, Peru
[200] Istituto Nazionale di Fisica Nucleare, Sezione di Torino, Turin, Italy
[201] Centro de Tecnologia da Informacao Renato Archer, Amarais–Campinas CEP 13069-901, SP, Brazil
[202] Gran Sasso Science Institute, L'Aquila, Italy
[203] University of Arizona, Tucson, AZ 85721, USA
[204] Punjab Agricultural University, Ludhiana 141004, India
[205] Jeonbuk National University, Jeonrabuk-do 54896, Republic of Korea
[206] Central University of South Bihar, Gaya 824236, India
[207] Universidad Nacional Mayor de San Marcos, Lima, Peru
[208] Institute of Particle and Nuclear Physics of the Faculty of Mathematics and Physics of the Charles University, 180 00 Prague 8, Czech Republic
[209] National Institute of Science Education and Research (NISER), Odisha 752050, India
[210] Texas A&M University, Corpus Christi, TX 78412, USA
[211] University of Medellín, Medellín 050026, Colombia
[212] Idaho State University, Pocatello, ID 83209, USA
[213] Jyväskylä University, FI-40014 Jyväskylä, Finland
[214] Kavli Institute for the Physics and Mathematics of the Universe, Kashiwa, Chiba 277-8583, Japan
[215] Laboratori Nazionali del Gran Sasso, L'Aquila, AQ, Italy
[216] University of California Los Angeles, Los Angeles, CA 90095, USA
[217] Institute of High Energy Physics, Chinese Academy of Sciences, Beijing, China
* Correspondence: ajwhite@uchicago.edu

**Abstract**

The 2x2 Demonstrator, a prototype for the Deep Underground Neutrino Experiment (DUNE) liquid argon (LAr) Near Detector, was exposed to the Neutrinos from the Main Injector (NuMI) neutrino beam at Fermi National Accelerator Laboratory (Fermilab). This detector is a prototype of a new modular design for a liquid argon time-projection chamber (LArTPC), comprising a two-by-two array of four modules, each further segmented into two optically isolated LArTPCs. The 2x2 Demonstrator features a number of pioneering technologies, including a low-profile resistive field shell to establish drift fields, native 3D ionization pixelated imaging, and a high-coverage dielectric light readout system. The 2.4-tonne active mass detector is flanked upstream and downstream by supplemental solid-scintillator tracking planes, repurposed from the MINERvA experiment, which track ionizing particles exiting the argon volume. The antineutrino beam data collected by the detector over a 4.5 day period in 2024 include over 30,000 neutrino interactions in the LAr active volume—the first neutrino interactions reported by a DUNE detector prototype. During its physics-quality run, the 2x2 Demonstrator operated at a nominal drift field of 500 V/cm and maintained good LAr purity, with a stable electron lifetime of approximately 1.25 ms. This paper describes the detector and supporting systems, summarizes the installation and commissioning, and presents the initial validation of collected NuMI beam and off-beam self-triggers. In addition, it highlights observed interactions in the detector volume, including candidate muon antineutrino events.

## 1. Introduction

The Deep Underground Neutrino Experiment (DUNE) [1] is an accelerator-based neutrino observatory currently under construction along the new Long-Baseline Neutrino Facility (LBNF) neutrino beam [2]. DUNE will unambiguously determine the neutrino mass ordering; it will also measure the charge-parity (CP)-violating phase and several neutrino mixing parameters with high precision. Additionally, DUNE will search for physics beyond the Standard Model and utilize any supernova neutrino bursts occurring in its lifetime, close enough to be detected, to study the astrophysics of stellar collapse and the properties of neutrinos [3]. DUNE is composed of a Near Detector (ND) complex located at Fermilab in Batavia, Illinois, and a Far Detector (FD) complex positioned 1285 km away at Sanford Lab in Lead, South Dakota. More information about the experiment and its physics goals can be found in [1,4].

The FD complex will be constructed in two phases, with Phase I consisting of two 10 kilo-tonne (counting fiducial mass within the instrumented regions, with each cryostat holding a total of 17.1 kt LAr [5]) LArTPCs and an addition of two more detectors available in Phase II. The ND complex [5] will house three detectors, all situated 574 m downstream of the LBNF neutrino source in a cavern with approximately 60 m of rock overburden. To minimize systemic uncertainties for the oscillation analyses, it is essential to include an LAr target in the ND complex. ND-LAr, an LArTPC-based Near Detector, satisfies this requirement. Together with The Muon Spectrometer (TMS), a magnetized scintillator detector located directly downstream along the beamline, ND-LAr will have the capability of moving orthogonally to the beam axis in order to characterize the LBNF neutrino flux over a range of angles, a technique referred to as PRISM [5]. The third detector in the ND complex, System for on-Axis Neutrino Detection (SAND), will remain fixed to characterize and monitor the on-axis beam continuously.

The LBNF beam, currently under construction and slated to become the highest-intensity neutrino beam in the world, will present new challenges for LArTPC neutrino detectors. During Phase I, with LBNF in forward horn current mode, ND-LAr expects an average of 25 neutrino interactions on argon alongside O(100) background events per 9.6 µs beam spill; however, its detector design must also accommodate an increased neutrino interaction rate following the planned Phase II LBNF beam upgrades. To meet DUNE physics requirements, the ND-LAr design has been developed to mitigate interaction pileup in its uniquely high signal-occupancy environment. It features a pixelated charge readout that produces 3D ionization distributions to which reconstruction algorithms can be directly applied; this charge readout works in tandem with a fast (nanosecond-level), high-coverage light readout system within a modular detector design. The modularity of ND-LAr ensures the correct association of charge and light signals for accurate event reconstruction and energy deposition estimation.

A staged prototyping program has been implemented to facilitate the success of the innovative ND-LAr design. The 2x2 Demonstrator represents the first multi-module stage of testing for the fully integrated system, consisting of four prototype modules running in the high-occupancy NuMI beam at Fermilab [6,7]. Each module is approximately 60% of the full-scale design in each transverse dimension and 40% of the height compared to the planned ND-LAr modules. Prior to their installation in the 2x2 Demonstrator, each module was tested individually at the University of Bern in Switzerland [8]. The 2x2 Demonstrator sees a neutrino interaction rate per TPC that is comparable to that of ND-LAr, providing a valuable test of the novel detector geometries and technologies devised to mitigate event pileup.

This paper covers the details of, and motivation for, the 2x2 Demonstrator system's design in Section 2. Section 3 describes the procedures for the installation and commissioning of the detector and cryogenic systems. Section 4 reports on the initial detector performance based on raw data; it includes beam and self-triggering timing validations, as well as visually identified, triggered events.

## 2. Design of the 2x2 Demonstrator

This section will summarize the considerations and challenges driving the 2x2 Demonstrator design. It will also describe the detector system layout as a whole, including cryogenic support systems and the NuMI beam configuration.

### 2.1. LArTPC Detectors: Advantages and Challenges

Since their first large-scale implementation in the early 2000s [9], LArTPCs have furnished the physics community with high-granularity imaging and precise calorimetry in a dense, scalable, uniform, and fully active detector medium. In conjunction with its high density ($\rho$ = 1.4 g/cm$^3$), LAr provides high ionization and scintillation yields of $\mathcal{O}(10^4)$ photons and electrons per deposited MeV [10,11], enabling excellent ionizing track reconstruction and making it an ideal target material for neutrino detectors.

LArTPC particle detectors such as ArgoNeuT [12], LArIAT [13], ICARUS [14], MicroBooNE [15], SBND [16], and ProtoDUNE-SP [17] have demonstrated that in a volume of liquid argon sufficiently free of electronegative impurities, ionization electrons can be drifted over the order of meters and imaged at mm scale resolutions [18–22]. The resultant scalability of the LArTPC increases the likelihood of fully containing neutrino interactions within a detector's fiducial volume, improving its calorimetric performance.

Within an LArTPC, energetic, charged particles produce argon excimers, which in turn decay and produce scintillation light; the fast (singlet state) component of this signal, propagating to the boundaries of a detector within nanoseconds of an interaction, provides an efficient $t_0$ timestamp. The slow (triplet state) component of the scintillation light follows on the order of a µs. In an environment such as the DUNE ND, with tens of neutrino interactions spanning a 9.6 µs window, even the fast components of scintillation signals can experience pileup if multiple signals become indistinguishable in time. Particularly for interactions with final-state particle tracks detached from the interaction vertex, pileup of fast-component light signal risks merging neutrino interactions during event reconstruction, which can bias calorimetric measurements and event classification.

### 2.2. ND-LAr: Pileup Mitigation

The ND-LAr design emerged from the ArgonCube development program [23], which aimed to improve upon the performance and robustness of the wire-plane LArTPC design through new and developing technologies. Specifically, ND-LAr seeks to minimize pileup-related biases in event reconstruction through TPC modularity.

Precise energy reconstruction and interaction-type classification will depend upon the successful separation of simultaneous events, a goal which requires ND-LAr to diverge from the monolithic LArTPC design and its traditional technologies. In an LArTPC with a maximum drift length of two meters, at a drift field of 500 V/cm corresponding to an approximate drift velocity of 0.16 cm/µs, the maximum drift time will exceed the length of a beam spill by more than 1000 µs. This difference in scale between the charge readout and spill windows (over two orders of magnitude) can lead to pileup of ionization tracks in a multi-interaction beam spill, an effect which scales with a detector's maximum drift length. To mitigate this pileup, rather than constructing one or two monolithic volumes, ND-LAr will be composed of 35 modules in a 7 × 5 × 3 m volume. Each module, measuring 1 × 1 × 3 m, will be split along the beam axis into two optically isolated TPCs, resulting in a total of 70 TPCs. The 2x2 Demonstrator prototypes this with four modules, totaling eight optically isolated TPCs.

ND-LAr modules will collect charge on pixelated sensors, a charge readout technique new to large-scale LArTPCs [24]. The two spatial dimensions defined by the pixelated anode, when combined with the relative timing of pixel hits, provide native 3D imaging of particle tracks within the detector volume before any reconstruction is applied. To complete this structure while also minimizing uninstrumented regions between each TPC, a novel, low-profile field shell and a dielectric light readout system (LRS) have also been designed [25,26]. Segmenting the detector into multiple independently instrumented, optically isolated regions and including a light readout to supply interaction timing on a scale capable of resolving the beam structure will improve the ability of ND-LAr to accurately reconstruct tens of neutrino interactions within individual beam spills, in spite of backgrounds and the expected pileup rate. The introduction of a small amount of dead space between these modular regions has no effect on the underlying interaction physics within each TPC, and its impact on reconstruction efficiency can be quantified and corrected for using data-driven calibration techniques. Taken together, the innovations implemented in ND-LAr, which are necessary to disentangle and reconstruct the very large number of interactions present in each beam spill, are expected to provide equivalent or improved reconstruction performance relative to the Far Detectors and preserve the validity of Near–Far comparisons.

In the DUNE ND complex, TMS is located downstream of ND-LAr to track and characterize muons exiting ND-LAr. For the 2x2 Demonstrator, scintillator planes interleaved with steel panels repurposed from the MINERvA [27] experiment (Mx2) flanking the LArTPC both upstream and downstream along the beam axis provide similar muon tracking and calorimetry. The Mx2 is described in greater detail in Section 2.7. To reconstruct partially contained events, precise timing is required to match energy depositions in ND-LAr to associated depositions in external detectors like TMS. This is especially true in environments with high rates of interaction pileup, when multiple overlapping or proximal particle trajectories could exit the LArTPC in the span of several microseconds. Event and interaction-level matching between the 2x2 Demonstrator LArTPC and the Mx2 will provide a valuable test of this process.

### 2.3. 2x2 Demonstrator: Overview

The 2x2 Demonstrator was installed 102 m underground in Fermilab's MINOS Hall, currently home to the NOvA ND [28] and previously home to the MINOS ND [29], ArgoNeuT [12], and MINERvA experiments. The 2x2 Demonstrator is centered on the beam axis, in the same location that the MINOS Near Detector once occupied; see Figure 1.

The four 60% scale prototype ND-LAr modules that make up the 2x2 Demonstrator are suspended in a common bath of liquid argon within a 6.1 $m^3$ cylindrical, vacuum-jacket-insulated cryostat. Together, the four modules form a 2.4-tonne modular LArTPC, pictured in Figure 2.

Each module hangs beneath a stainless steel top flange containing five feedthroughs: the detector power and readout cables for each TPC, the high voltage input, detector monitoring and light system calibration inputs, and an LAr fill port. Before insertion into the cryostat, the four modules are connected via a steel cross bar enforcing mm scale separation between modules. Indium was used to provide a cryogenic seal between the modules, crossbar, and cryostat top plate. The four module units can be seen in Figure 2 prior to their insertion into the cryostat.

Within each module, the two optically isolated TPCs are instrumented with pixel-based charge readout system (CRS) anodes positioned on either side of a central cathode plane. The light detection systems are located normal to the anode planes on both sides of each TPC; they consist of dielectric, scintillating light traps which provide 29% geometrical coverage within each module.

Located 0.47 m downstream and 0.75 m upstream from the cryostat are scintillator-based tracking detectors, which were repurposed from the MINERvA experiment.

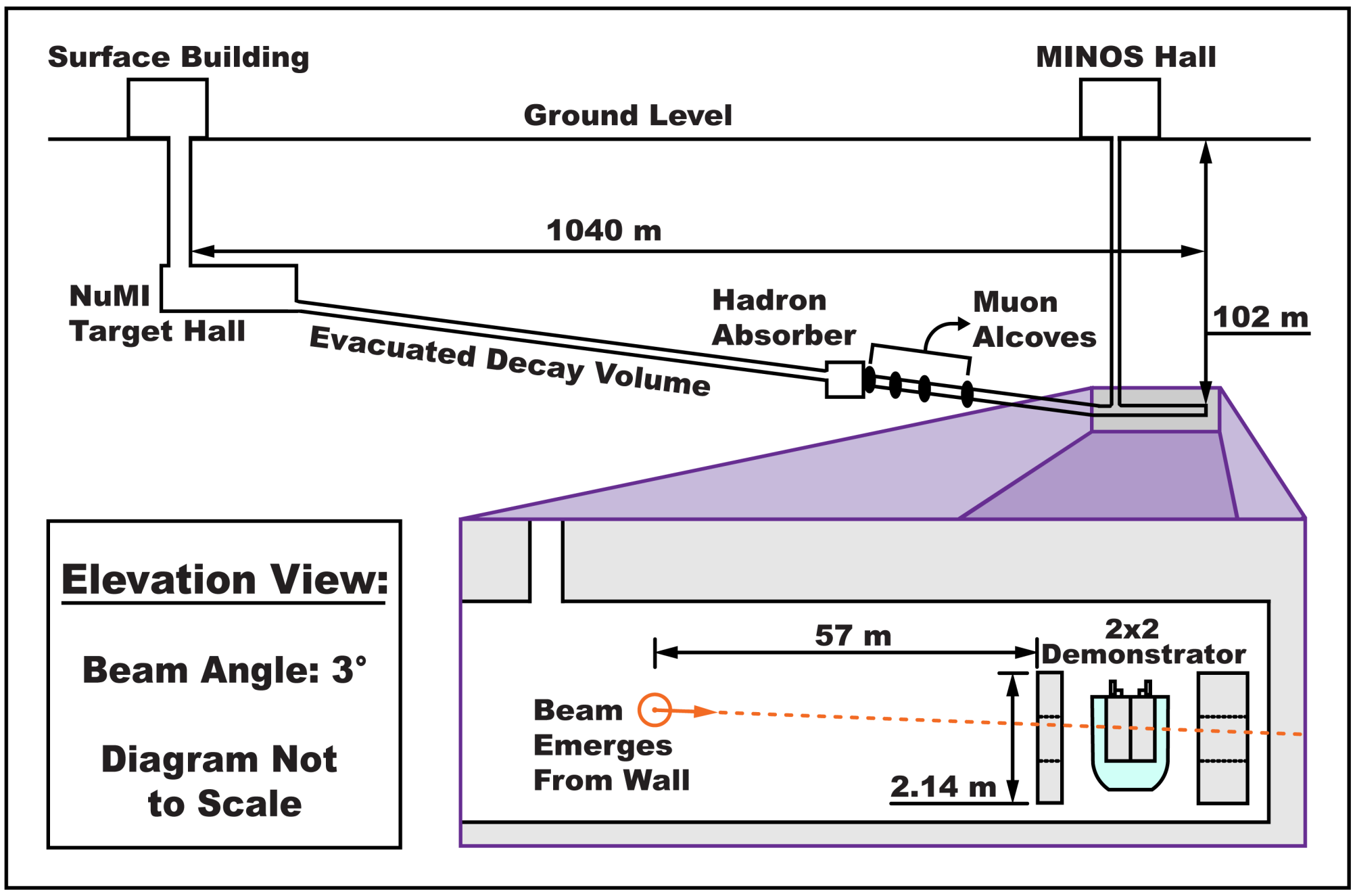

**Figure 1.** The NuMI Target Hall and Minos Hall; see inset. The 2x2 Demonstrator is located at the downstream end of the MINOS Hall.

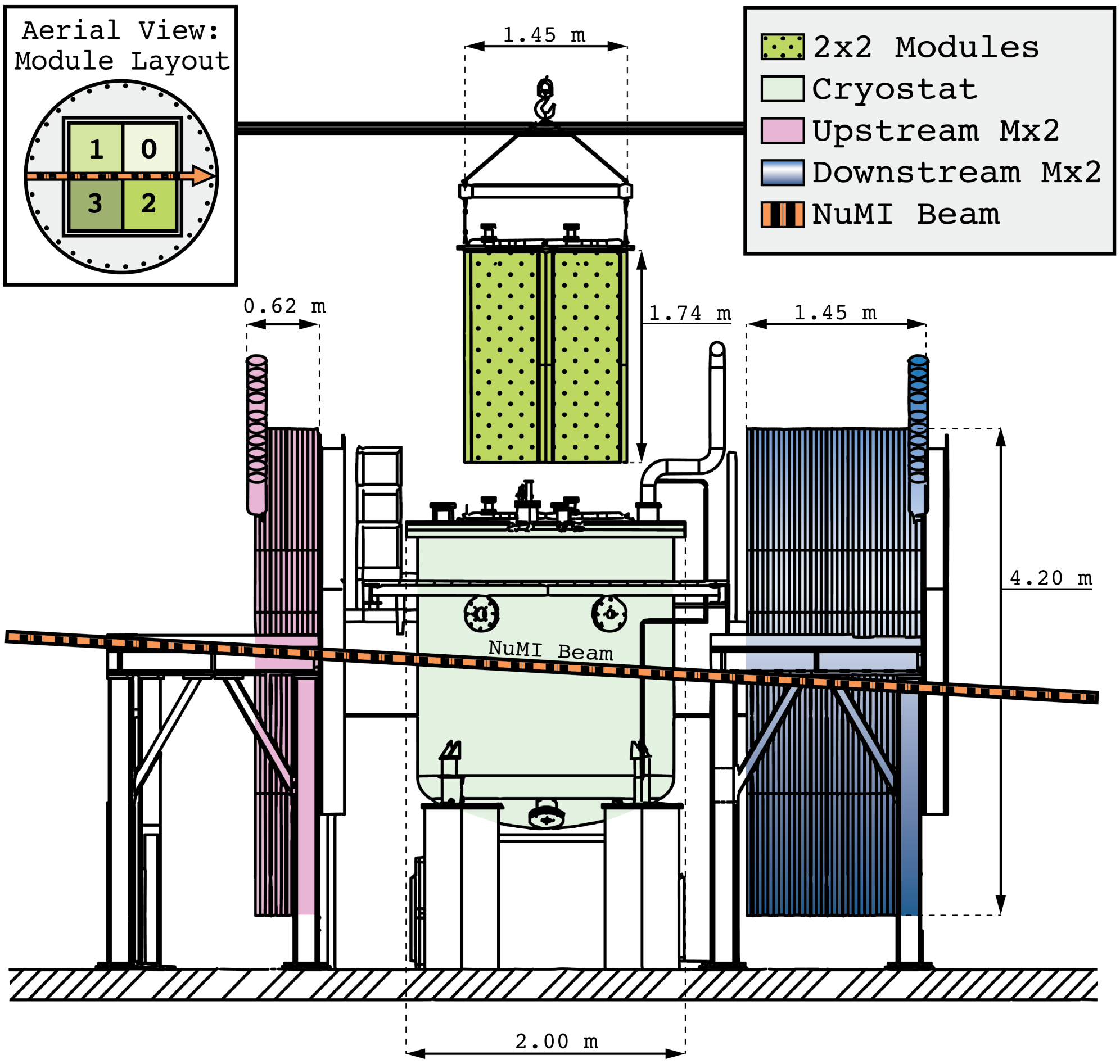

**Figure 2.** A rendering of the 2x2 Demonstrator and associated systems. The central cryostat contains four independent LArTPC modules in a common bath of liquid argon. Hexagonal steel panels interleaved with scintillator tracking planes, repurposed from the MINERvA detector, are located on both sides of the cryostat, along the axis of the incoming neutrino beam entering from the left.

### 2.4. Cryogenic System: Design and Monitoring

Due to the underground location of the 2x2 Demonstrator, direct filling from LAr or $N_2$ tankers was unfeasible. Consequently, the detector was filled via 160 L dewars lowered by crane into the MINOS Cavern. Throughout operations, LAr purity was maintained through recirculation, with an internal cryogenic centrifugal pump [8] extracting LAr from the bottom of the cryostat, passing it through $O_2$ and $H_2O$ filtering media [30,31], and pumping the LAr back into the cryostat through the fill ports positioned at the top of each module. Cooling was provided by three cryocoolers [32], driving cold heads in an LAr condenser. After filling the cryostat, a non-negligible leak was detected along the indium seal joining the modules to the cryostat top plate. A continuous top-up with gaseous argon mitigated the effects of this leak, with $H_2O$ and $O_2$ gas-getters installed on the input line to preserve LAr purity. Gas analyzers continuously sampled $O_2$, $N_2$, and $H_2O$ at various locations, including the input and output of the LAr filter.

The cryogenic system was designed to provide 40 mm of dielectric shielding of LAr to the high-voltage system. Due to a difference between the as-built detector and the design, the position of the condenser boil-off line limited the maximum dielectric shielding to 20 mm for the duration of the 2024 run. The maintained LAr level served as adequate shielding for the design voltage of 500 V/cm. Following the initial data-acquisition period in July 2024, the condenser boil-off line was moved, extending the possible voltage range for future runs to 1 kV/cm. A calibrated liquid-level sensor provided input to the high-voltage interlock whenever voltage was applied.

### 2.5. NuMI Beam

The 2x2 Demonstrator sits 1.04 km downstream of the NuMI target facility at Fermilab. To produce the NuMI beam, 120 GeV protons from Fermilab's Main Injector Accelerator are directed at a 1.2 m graphite target, yielding charged kaons and pions that are subsequently either focused or deflected from the beam path by electromagnetic focusing horns. The beam can run in both neutrino and antineutrino mode, depending upon the polarity of the current supplied to the focusing horns. For the duration of the 2024 run, NuMI ran with reverse horn current (RHC) to produce a muon antineutrino beam.

After the focusing horns, the hadrons produced at the beam target enter a 675 m decay pipe, followed by 240 m of rock. The muon antineutrinos emitted as the hadrons decay in flight have an average energy of 5.8 GeV (see Figure 3). For the medium-energy RHC configuration, the NuMI beam is expected to have a high purity of approximately 95% muon antineutrinos in the energy range of interest, which is 0 to 10 GeV. Muon neutrinos account for 4% of the background contamination, while electron (anti)neutrinos constitute the remaining 1%. Figure 1 shows the layout of the beam and injector facilities with respect to the MINOS Hall [6].

The Main Injector beam spill spans 9.6 µs, with varying cycle times depending on the configuration of the accelerator complex. Between 8 July and 12 July, when the 2x2 collected physics data, the cycle time between consecutive spills was approximately 1.2 s.

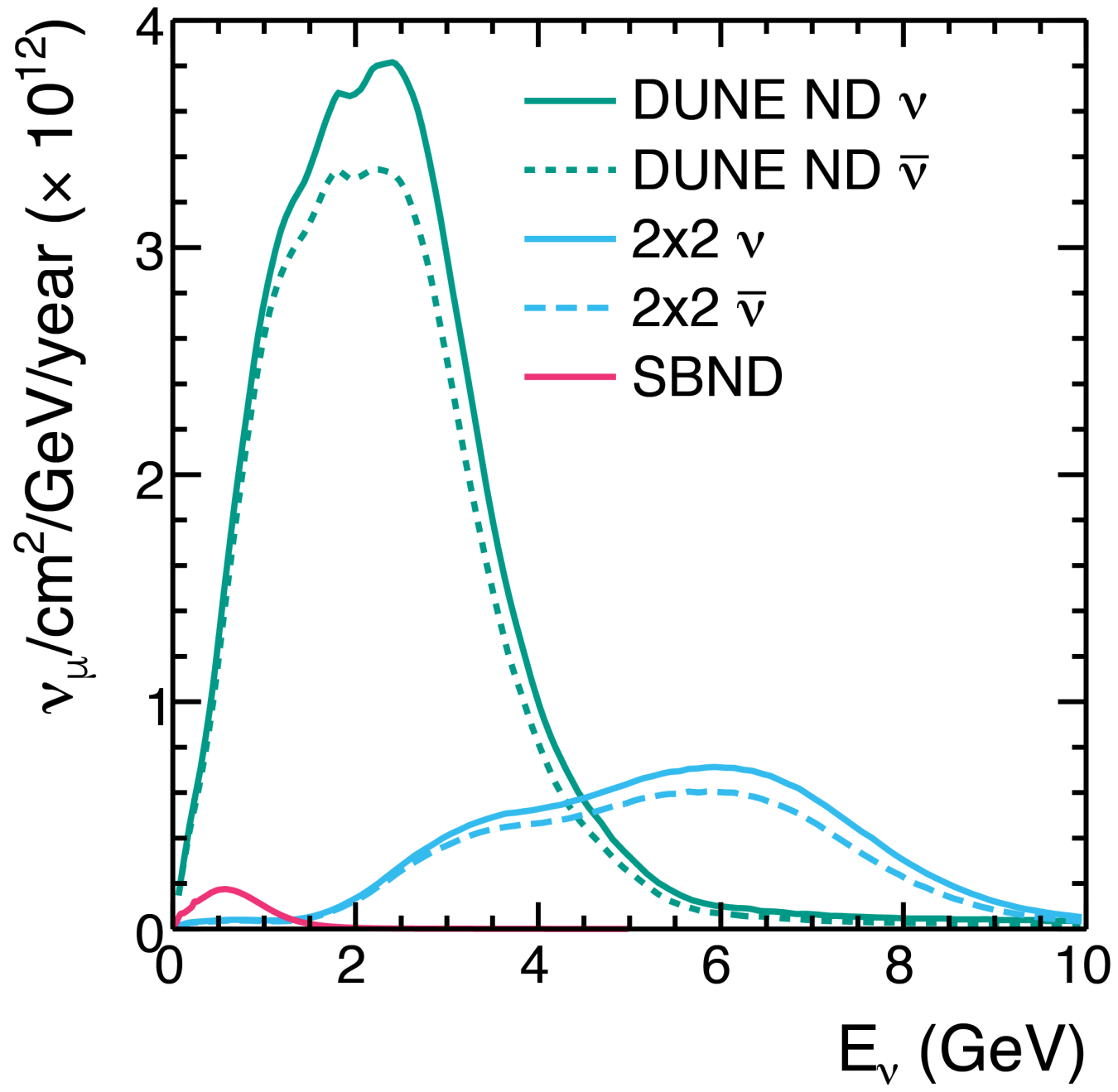

**Figure 3.** The expected energy and flux of neutrinos from the NuMI beam at the 2x2 Demonstrator, compared to those from the Booster Neutrino Beam (BNB) at the SBN Near Detector (SBND) [16] and the Phase I LBNF beam at the DUNE ND.

### *2.6. LArTPC Design and Readout*

Details of the 2x2 module design are summarized below, and a full description of module hardware can be found in [8]. (This is a report on the design and performance of the first 2x2 module, Module 0, built and operated at the University of Bern). As the four 2x2 modules were constructed and tested sequentially over several years, hardware upgrades and variations were integrated into the modules. The differences are summarized in Table 1. Modules will be referred to as Module 0, 1, 2, and 3, based on the order in which they were constructed.

**Table 1.** Aspects of the readout subsystems that differ across modules: all modules use 8 tiles per anode.

| Feature | Module 0 | Module 1 | Module 2 | Module 3 |
|---|---|---|---|---|
| Pixel Pitch [mm] [a] | 4.43 | 4.43 | 3.88 | 4.43 |
| Pixels/Tile | 4900 | 4900 | 6400 | 4900 |
| SiPM Model [b] | 6025PE | 6050PE | 6050PE | 6050PE |
| Edge Dichroic Mirror [ACL] | No | Yes | Yes | Yes |

[a] Distance between the centers of two adjacent pixels; [b] Hamamatsu S13360 (Hamamatsu, Japan).

The 2x2 modules are housed in 1.4 m tall sleeves of 6 mm thick G10/FR-4 [33], each with a 0.7 m × 0.7 m footprint. The sleeve's base is left open to facilitate LAr circulation and its top is attached to the cryostat flange [5]. The choice of G10/FR-4 is advantageous as a construction material as it has a radiation length of 19.4 cm and a hadronic interaction length of 53.1 cm, both of which are on the same scale as LAr (14 cm and 85.7 cm, respectively) [34]. In comparison, iron has radiation and hadronic interaction lengths of 1.75 and 16.7 cm, respectively, so a thin G10 shell behaves far more like LAr than a traditional stainless steel field cage [35]. This similarity reduces potential event reconstruction complications introduced by the uninstrumented material between fiducial volumes.

2.6.1. LArTPC Subsystem: Drift High Voltage

Each 2x2 module is split along the beam axis into two optically isolated TPC volumes sharing a central cathode. The maximum drift distance from anode to cathode across each TPC measures 30 cm, requiring only a modest −15 kV potential on the cathode to achieve the nominal field strength of 500 V/cm. This drift high voltage (HV) is supplied for all modules by a Spellman [36] SL50N300/ESL/220 unit. A custom-built oil-filled potted filter-distributor [37] splits the drift HV into four output channels and provides resistive decoupling for the four, while also acting as a low-pass filter for the power supply ripple.

The uniformity of the electric field within each TPC is enforced by a resistive field shell—a departure from the traditional LArTPC resistor-chain field cage. The resistive field shell, constructed from a layer of carbon-loaded Kapton film (DR8) laminated to G10, surrounds each drift region as pictured in Figure 4. At a field strength of 500 V/cm, the 100 μm film has a sheet resistance of $\mathcal{O}(10^9)$ Ω/sq [34]. The field shell is coupled to the anode and cathode by copper cladding; each cathode panel, also constructed from G10/FR-4 and carbon-loaded Kapton film (XC, 25 μm thick), has a reduced sheet resistance of $\mathcal{O}(10^6)$ Ω/sq. Cathodes feature Kapton film on both sides, as they bisect the modules [38]. This low-profile field shell design minimizes the uninstrumented material between the instrumented volumes.

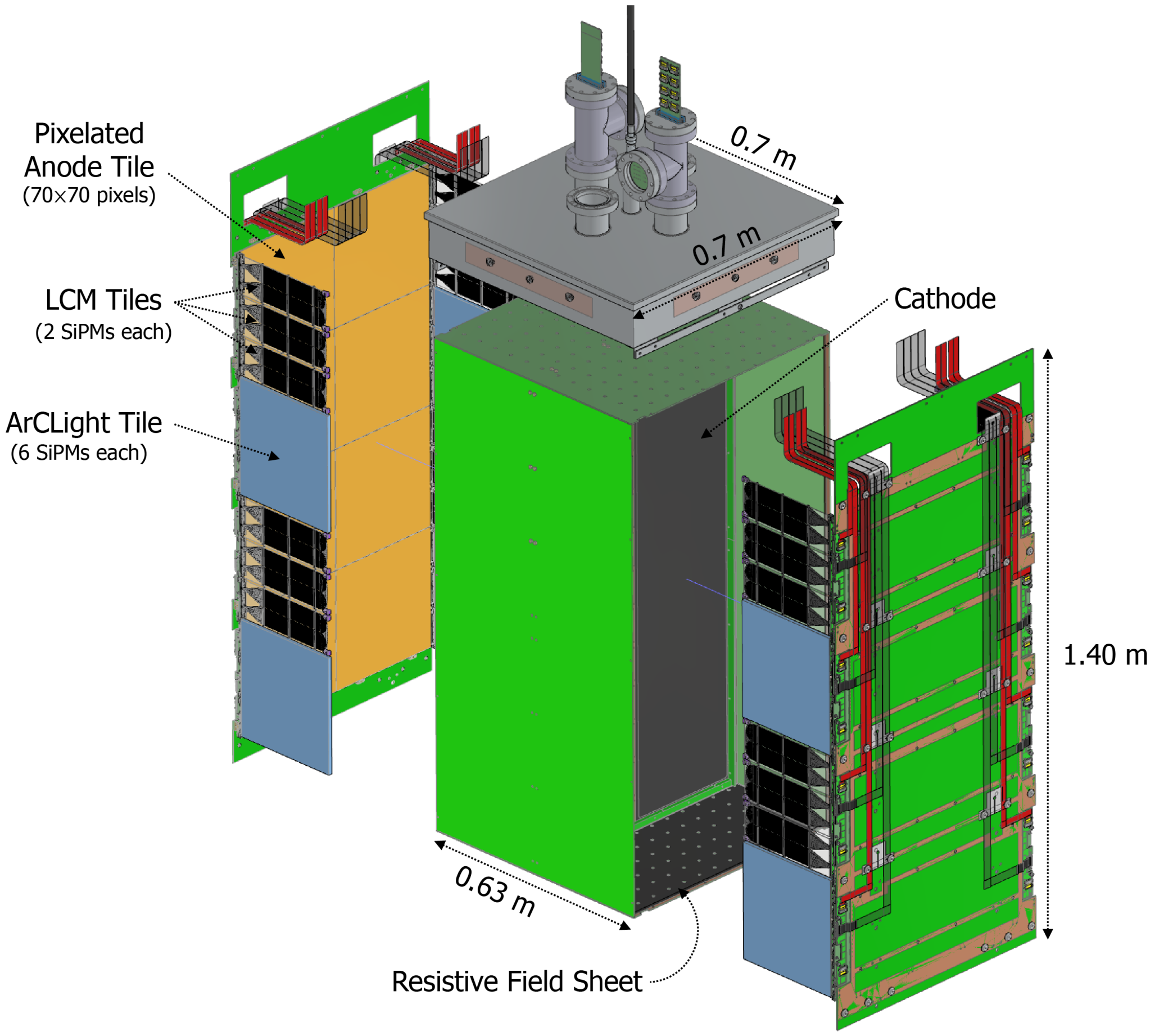

**Figure 4.** Rendering of a single 2x2 module, as described in [8]. The module is bisected by a resistive cathode to form two optically isolated LArTPCs. Readout cable positions for the CRS (gray) and LRS (red) can be clearly seen external to the anode. During operation, modules are protected by an external G10 sleeve.

#### 2.6.2. LArTPC Subsystem: Charge Readout

Positioned opposite to the central cathode in each module, two anode planes define the outer extent of the mirrored drift regions. Each anode is instrumented by eight large-format, low-profile, printed circuit board LArPix pixel tiles [24]. These are arranged in a 4-high-by-2-wide array, with each individual tile measuring 31 cm × 32 cm. The LArPix application-specific integrated circuits (ASICs) provide cryogenic-compatible, ultra-low-power amplification, self-triggered digitization, and digital multiplexing for a total of 337,600 pixels in the 2x2 Demonstrator [39]. Each ASIC controls 64 pixel channels, and is packaged to facilitate commercial production and mounted in an array to the back of the anode. Every pixel functions as an independent self-triggering detector and resets with a negligible (approximately 100 ns) dead time. 100 ns corresponds to an electron traveling 0.16 mm in the detector, given a nominal drift field of 500 V/cm. A typical pixel hit integrates charge for 2.6 μs, so two back-to-back hits on the same individual pixel separated by a reset would result in a loss of approximately 4% of the second track's charge. This effect impacts tracks traveling perpendicular to the anode, and can be suppressed by alternative LArPix modes. Consequently, the pixelated readout maintains a manageable data rate and well-resolved tracks even in a high-occupancy environment with multiple events. The raw hit data extracted from each pixel combined with the relative timing of hits across all triggered pixels provide native 3D imaging of particle footprints before any signal processing, filtering, or reconstruction is applied. Timing information from scintillation light detected by the LRS fixes the global position of charge hits in each TPC drift direction. Per TPC, a single on-detector Pixel Array Controller and Network card (PACMAN) unit delivers clean power and establishes I/O with the ASICs, sets configurations, and provides the data acquisition (DAQ) interface for all eight tiles. As mentioned previously, minor prototyping variations between modules are summarized in Table 1.

#### 2.6.3. LArTPC Subsystem: Light Readout

Along the inner face of the vertical field cage panels (normal to the anode plane), eight low-profile, dielectric scintillation light traps span the distance between each anode and the cathode on both sides of the TPC. As mentioned previously, these light traps provide 29% geometrical coverage within each module. Two complementary light trap types alternate along each TPC side, as shown in Figure 4; these are the Light Collection Modules (LCMs) [26] and the ArCLight (ACL) tiles [25,40].

The LCM consists of wavelength-shifting (WLS) fibers coated in tetraphenyl butadiene (TPB) and coupled at each end to silicon photomultipliers [41] (SiPMs). Two SiPMs per LCM are mounted side-by-side with their associated electronics to a 30 cm × 10 cm × 0.1 cm PVC backing plate; scintillating fibers are bent, laid flat, and affixed to the same plate such that their ends couple securely to the SiPMs. Within the detector, LCMs are installed in sets of three and cumulatively occupy an area comparable to that of one ACL. The ACL is a 1 cm thick plate of WLS plastic covered by a TPB-coated dichroic mirror foil. In all modules except Module 0, additional strips of dichroic foil prevent light from escaping along the edges of the ACL. A single ACL tile measures 28 cm × 30 cm × 1 cm. ACLs are coupled to six SiPMs each [41], for a total of 384 SiPMs in the 2x2 Demonstrator. Schematics of both light trap types are provided in Figure 5.

Signals from the SiPMs undergo pre-amplification on cold electronics, while external variable-gain amplifiers and analog-to-digital converters (ADCs) sample and record the signals at 62.5 MHz with 14-bit resolution. The LRS and CRS both receive a trigger issued in coincidence with the neutrino beam pulse, prompting on-beam readout. Off-beam, every set of six SiPMs triggers independently on an adjustable threshold in order to detect activity like cosmic rays. Blue LEDs located within each TPC, controlled by a custom pulser unit, provide an internal gain calibration source for the individual SiPMs.

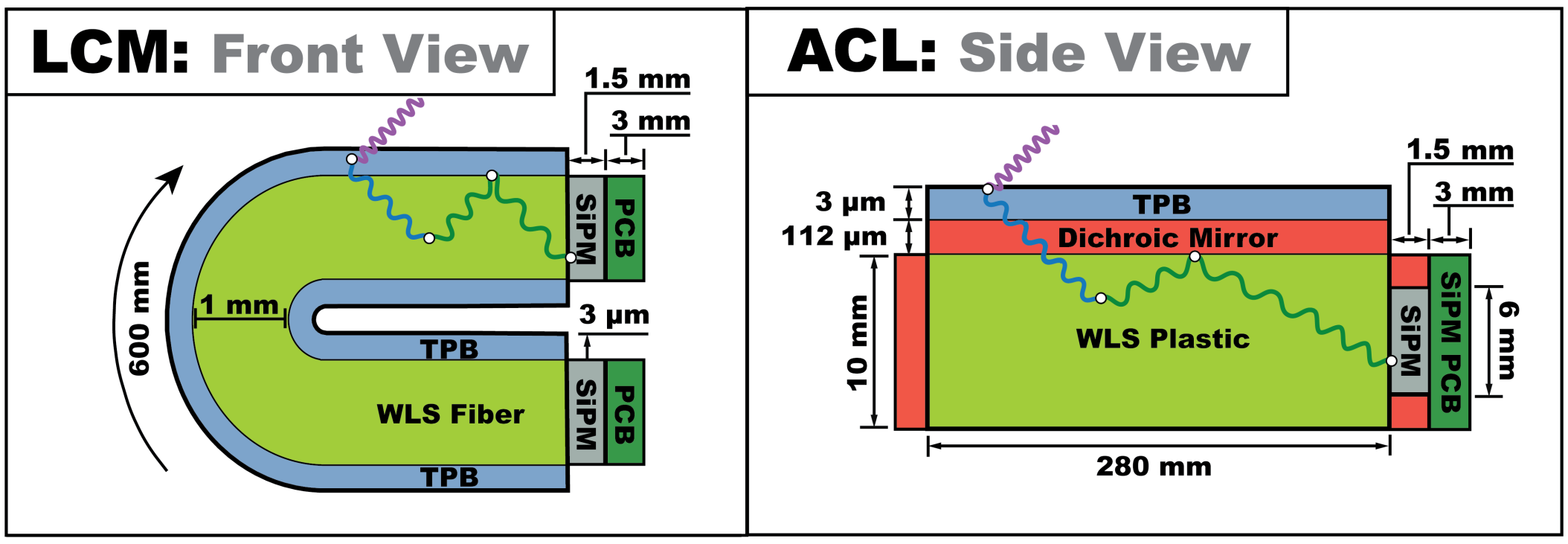

**Figure 5.** Schematics of the two LRS light trap types deployed in the 2x2 Demonstrator: the LCM (**left**) and ACL (**right**) with their respective dimensions. Installed in the detector, individual LCMs measure 10 cm vertically, while a single ACL spans 30 cm along the same dimension. The oscillating lines represent possible photon trajectories, with interactions at each white point shifting the 128 nm scintillation light to a more easily detected 512 nm (green).

Both the ACL and LCM are constructed entirely of dielectric materials and can consequently be placed inside the detector volume without causing significant electric field distortions [8]; this placement optimizes light yield and minimizes the material separating instrumented LAr volumes. The LRS design provides a detailed timing structure with vertical spatial resolution. The aim of this design, prototyped in the 2x2 Demonstrator to meet the physics needs of ND-LAr, is to mitigate the neutrino interaction pileup expected in a near-beamline detector.

In the 2x2 Demonstrator, the TPC maximum drift length defines a charge readout window of approximately 200 μs, while the NuMI beam spill spans only 9.6 μs. As previously mentioned, this order of magnitude difference between the charge readout and spill windows can lead to the pileup of ionization tracks from different interactions occurring at different points within a single beam spill. The LRS helps to detangle this pileup: with ACL and LCM light detectors operating on either side of the anode, running along its full vertical length, the LRS achieves sufficient spatial resolution along the vertical and beam-direction axes to define regions of interest within each TPC. Each region is then assigned a timestamp, using timing granularity finer than the NuMI beam bunch structure. The LRS detects individual scintillation signals with nanosecond-level resolution; importantly, this resolution enables the association of daughter particles that are detached from their parent vertices, as in the case of neutral particle decays. By correctly associating ionization charge clusters with significant optical events centered in the same modular region, the ionization signals for each separate neutrino interaction within a spill can be identified and separated in spite of the 200 μs charge readout window.

### 2.7. Mx2: External Scintillator

Scintillator planes, repurposed from MINERvA, are positioned upstream and downstream of the 2x2 Demonstrator cryostat (see Figure 6), centered along the NuMI beamline. This supplemental system provides tracking for ionizing particles that pierce the LArTPC detector and are particularly useful for charged pion and muon discrimination. The planes are composed of WLS fibers encased in plastic scintillator strips, arranged in 3 orientations with respect to the beam for 3D reconstruction. Three different configurations of scintillator planes are used: tracker planes composed solely of scintillator strips, as well as electromagnetic and hadronic calorimeter planes, interleaved with lead sheets and steel plates, respectively. Figure 7 provides an overview of their composition, and a detailed description can be found in [27].

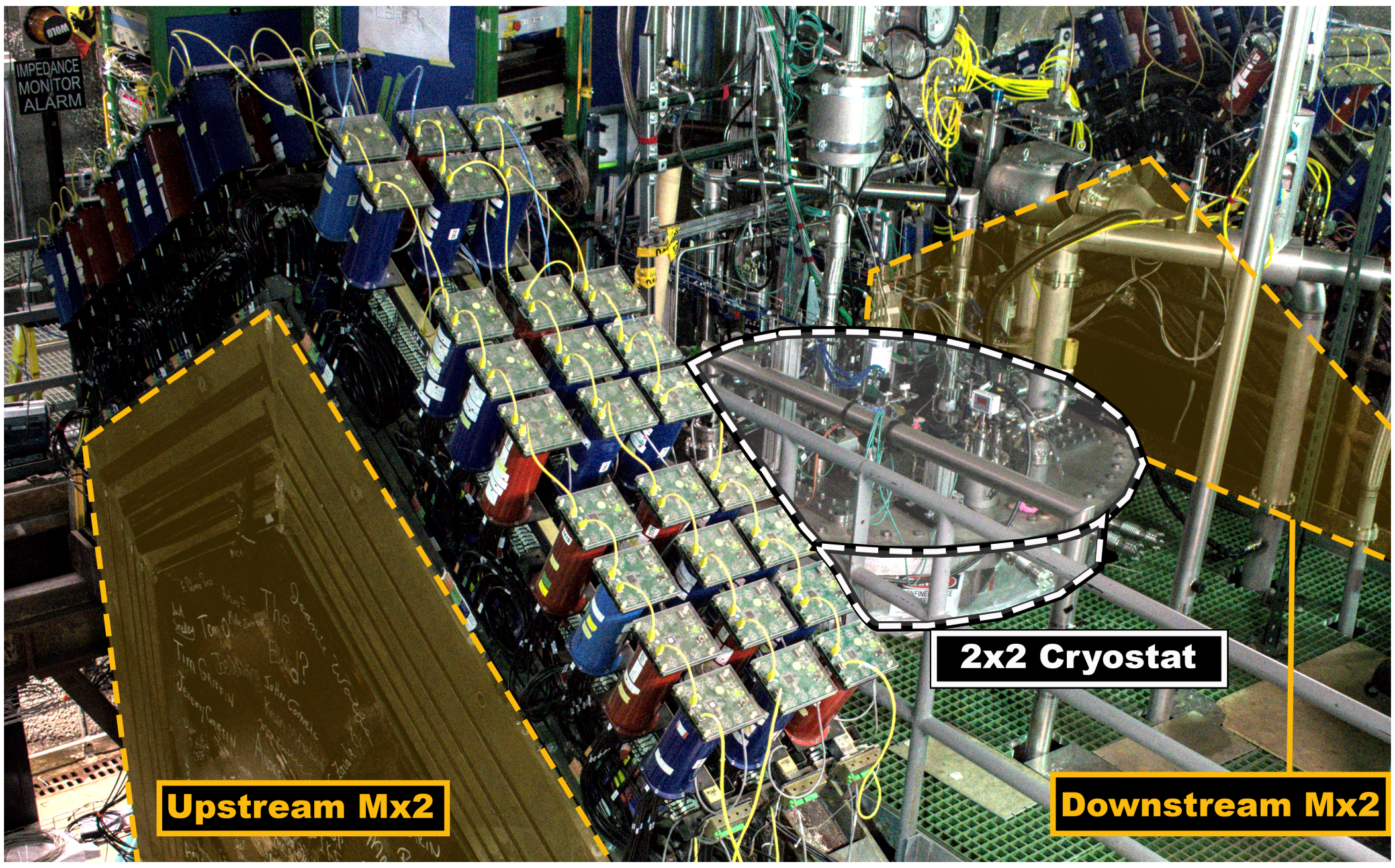

**Figure 6.** The 2x2 Demonstrator consists of a 2x2 array of modular LArTPCs flanked upstream and downstream by Mx2 scintillating tracker planes.

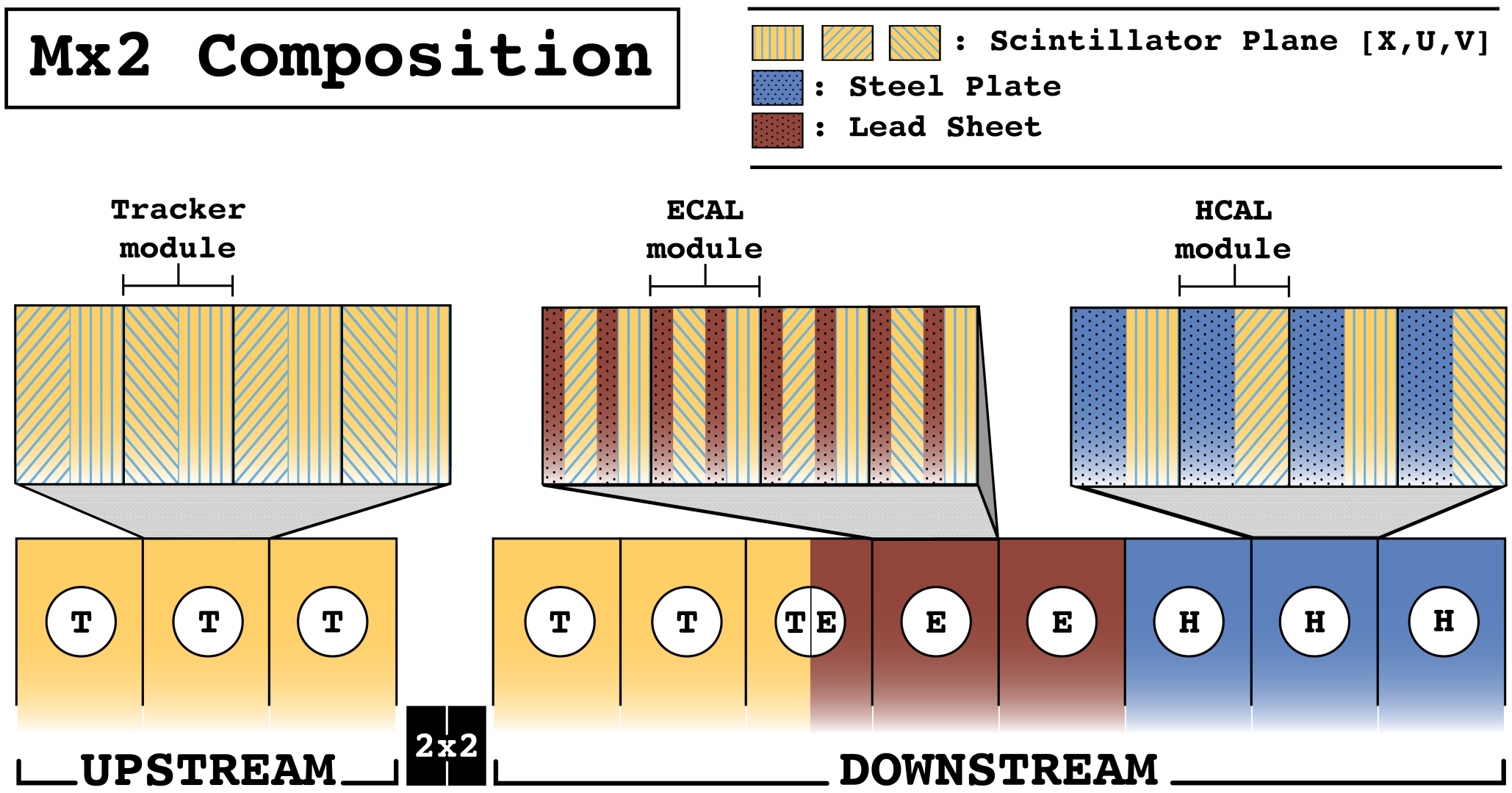

**Figure 7.** Muon taggers positioned 75 cm upstream and 47 cm downstream of the 2x2 LArTPC consist of three different types of tracking planes. The downstream modules were designed to also perform calorimetric measurements. A total of 185 128-channel multi-anode Hamamatsu H8804 PMTs are mounted atop the Mx2 module sets (one set consisting of four sequential modules); the upstream modules sport 57 PMTs, while the downstream modules operate the remaining 128.

Between the upstream and downstream components, the Mx2 supplies approximately 5.6 tonnes of active, fiducial scintillator mass, made up of 7.8% hydrogen and 92.2% carbon atoms. The Mx2 has a spatial resolution of 3.1 mm and a timing resolution of 3 ns [27]. The Mx2 reuses readout electronics from MINERvA, including original photomultiplier tubes (PMTs), front-end mother boards, and in-rack custom electronics and power supplies. The DAQ [42], as developed for MINERvA, was reused with minimal changes. The Mx2 DAQ triggers on the NuMI A9, a TTL (Transistor-Transistor Logic) early-warning signal indicating an incoming beam pulse. For the Mx2, receipt of the NuMI A9 initiates a 216 μs delay, after which a 16 μs readout gate is opened on all channels to capture the entire NuMI beam pulse as well as delayed light from decaying particles inside the detector.

## 3. Installation and Commissioning

The following section outlines the activities from the initial module construction and cosmic data acquisition at the University of Bern to the installation of the full 2x2 Demonstrator at Fermilab.This section also describes the commissioning of the 2x2 Demonstrator cryogenic and detector systems in the lead-up to its run in 2024.

### *3.1. Module Construction and Installation*

Between 2021 and 2023, the 2x2 Demonstrator modules were sequentially constructed and operated at the University of Bern, where they recorded cosmic ray muons in pure LAr for approximately one week per test run. Details of the assembly and performance of Module 0 were published in [8].

Over the same period, the infrastructure in the MINOS cavern at Fermilab was upgraded in preparation for the 2x2 Demonstrator installation. Both the MINOS and MINERvA experiments were removed from the cavern, while electrical, networking, and environmental safety systems were updated to meet the requirements of the 2x2 Demonstrator. In 2022 and 2023, the cryostat and the Mx2 were installed and their control systems subsequently commissioned.

As the modules completed their individual test runs at the University of Bern, they were shipped fully assembled to Fermilab and underwent warm acceptance tests on arrival.

In October 2023, the four modules were lowered 102 m by crane into the MINOS cavern. Once underground, they were arranged into a two-module by two-module array and installed in their cryostat, as shown in Figure 8. Installation of the remaining site infrastructure and supporting cryogenics system was completed in March 2024, whereupon the 2x2 Demonstrator subsystems collected warm commissioning data in the NuMI beam for detector calibration. In May 2024, purging of the 2x2 Demonstrator cryostat commenced, and the detector was filled with LAr by 31 May.

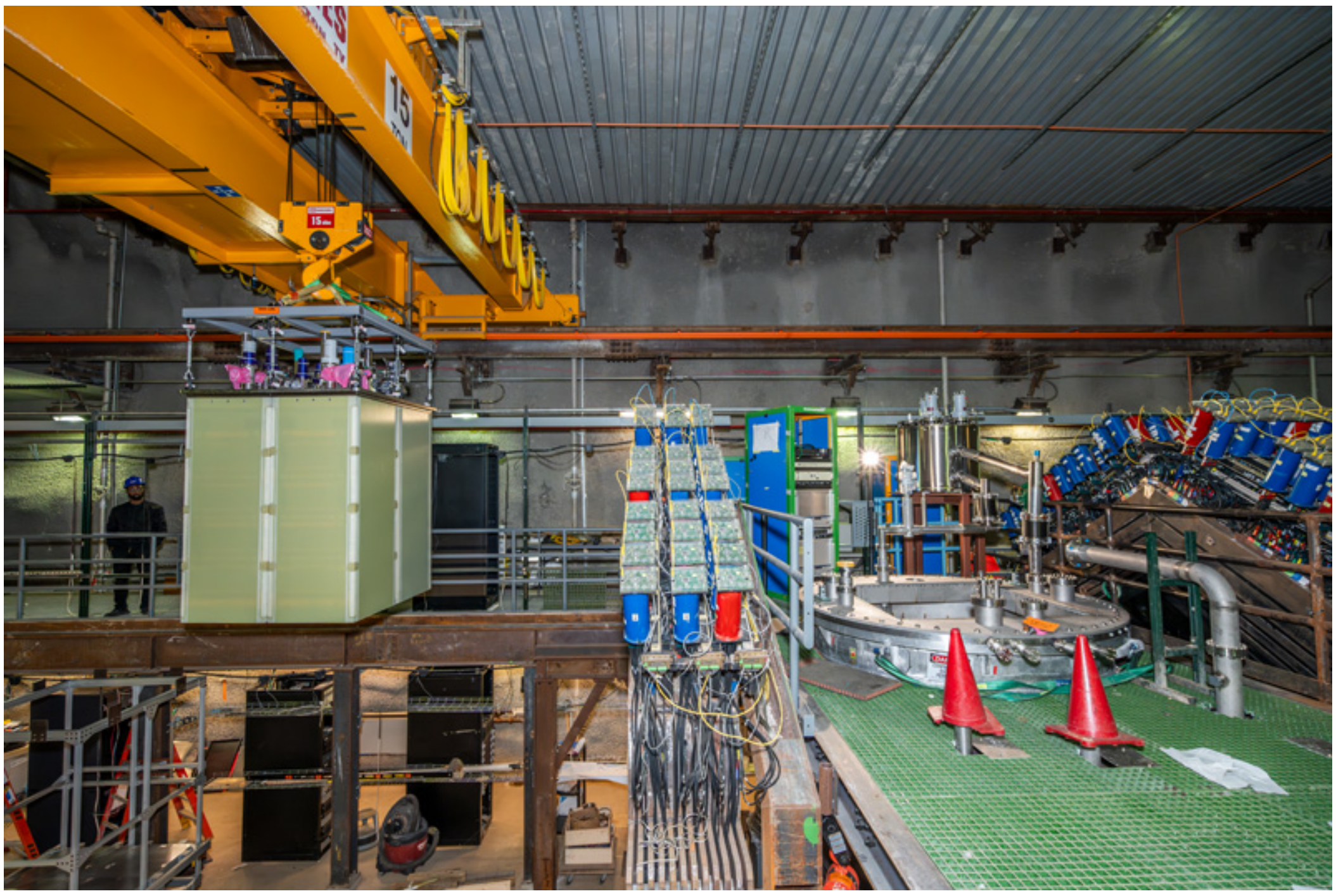

**Figure 8.** *Cont.*

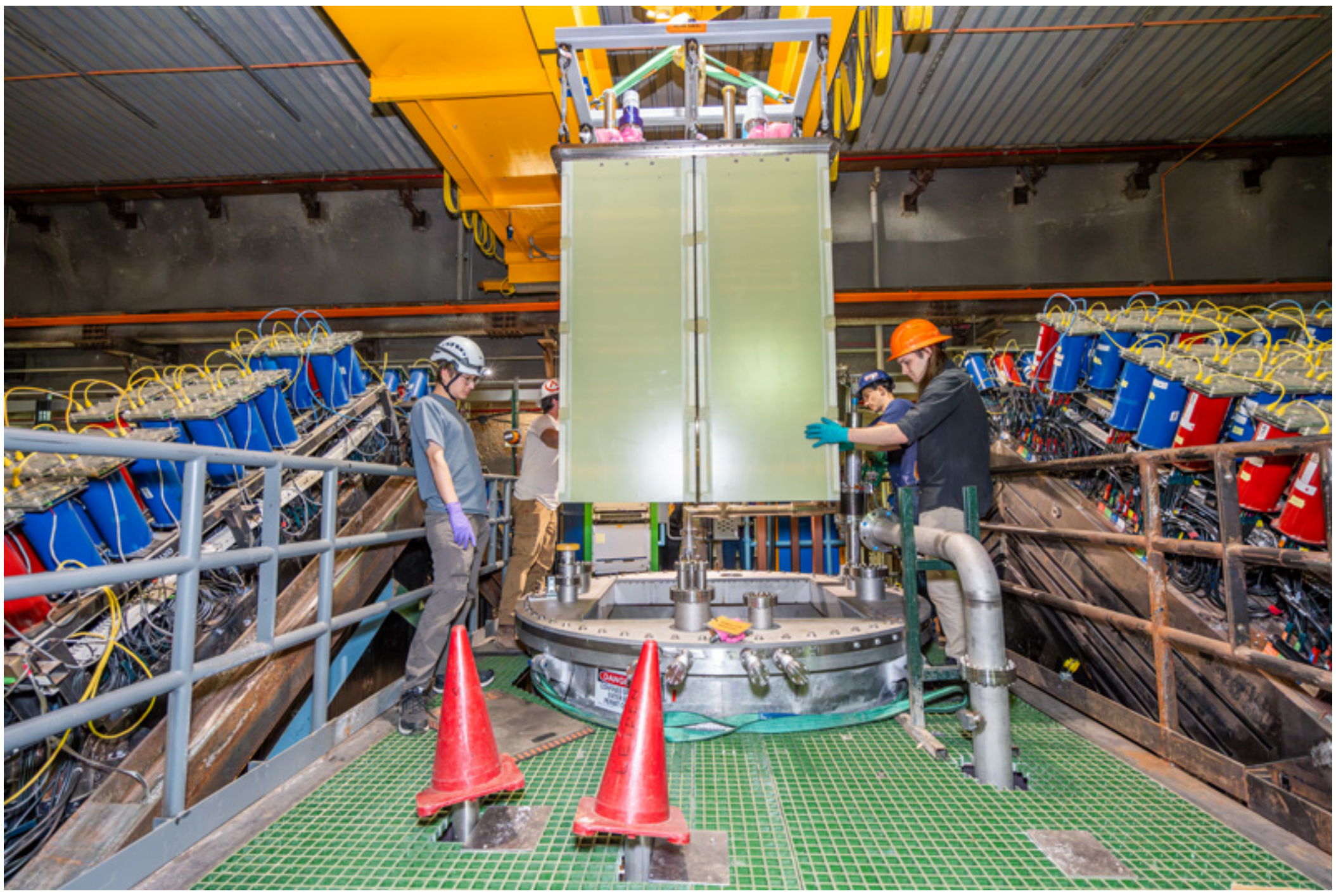

**Figure 8.** Installation of the 2x2 module array in the MINOS Underground Hall on 23 October 2023 [43].

### 3.2. Commissioning: Cryogenics

Leak-checking of the vessel after filling revealed a gaseous argon (GAr) leak in the indium seal joining the modules and the cryostat. This leak necessitated the installation of a continuous GAr top-up system, as well as $O_2$ and $H_2O$ gas purifiers to minimize added contaminants from the GAr. Additionally, a malfunctioning component of the detector condenser needed replacement, impacting the system's ability to maintain a high level of LAr purity. Once a spare was acquired, cryogenic commissioning continued. The cryogenic adjustments made in the period following the initial fill corresponded to an electron lifetime of approximately 100 μs, resulting in a charge loss of ∼86% for signals drifted over the full 30 cm anode-to-cathode distance during this initial period of operation.

Between 5 July and 6 July, in a brief window between high voltage runs, new cartridges were installed in the $O_2$ and $H_2O$ gas-getters. This maintenance period mitigated the full-drift charge attenuation to ∼15%, establishing an improved electron lifetime, which remained stable at approximately 1.25 ms for the remainder of the 2024 run.

### 3.3. Commissioning: TPC High Voltage

During cryogenic commissioning, several cross-checks were performed to verify the successful integration of the HV system. The resistances across the HV feed-throughs were measured, confirming good connections at the cathodes; the potted filter-distributor, system monitoring and controls, and the interlocks were likewise tested.

Cryogenic commissioning completed shortly before the Fermilab Summer Accelerator Shutdown in 2024; consequently, limited time was available for cold commissioning of the detector with the HV system fully ramped. The LRS provides HV monitoring during ramps, as potential electric breakdowns will produce flashes of light in the detector. The low cosmic muon rates underground improved the LRS sensitivity to any such flashes relative to the single-module runs at the University of Bern. A trial ramp to half-nominal (250 V/cm) field strength was performed on 31 May to commission the system after the detector was initially filled. During this test, slight deviations from nominal light rates were observed in the 2x2 Demonstrator. Offline investigations concluded that the elevated rates were due to low LAr purity rather than complications within the HV system.

On 1 July, the TPC HV system began ramping at a conservative rate of 25 V/s, pausing every 0.5 kV to monitor the status of the TPCs. The detector was held at half-nominal voltage for 12 h to ensure HV stability before completing its ramp to nominal voltage. The aggregate ramp from null to nominal TPC HV ($-15$ kV on the cathode) took 28.8 h.

On 4 July, an unintentional cessation of the HV occurred when the liquid-level interlock tripped, providing an opportunity to test the monitoring system's stability and robustness. The system and detector suffered no ill effects due to this HV trip, and cryogenic work leading up to the 7 July ramp significantly improved the LAr purity.

On 7 July and 8 July, the 2x2 Demonstrator was ramped to nominal voltage and ran continuously until midday on 12 July.

### *3.4. Commissioning: Light Readout*

Prior to the conclusion of cryogenic commissioning, the LRS underwent several commissioning stages. Under warm conditions, connections along the cold and warm readout chain were tested and confirmed in stages. At this preliminary stage, 13 SiPMs were flagged as non-working or exceeding acceptable noise levels, constituting 3.5% of all LRS channels. Once the cryostat was filled with LAr, the current draw of each cold electronics board was confirmed to lie within the expected range. Between the June 2024 and the July 2024 HV ramps, an ADC unit was replaced due to variable baselines, reducing the number of non-working or noise-dominated channels to 11 (2.8% of the total). The LED settings for each module were tuned for every SiPM and then deployed in automated calibration scripts. Calibration data were then taken at varying SiPM bias voltages, increasing the bias in steps of 0.5 V between each calibration run in order to determine optimal overvoltage settings for each SiPM. The average bias voltage across all operational LRS channels during the 2024 run was 46.8 V, corresponding to an average breakdown voltage of 42.3 V and an overvoltage of 4.5 V. Following this optimization of the LED and SiPM settings, calibration data were once again taken to extract the gain and resolution of each channel.

Due to the tight commissioning timeline driven by the beam shutdown, the opportunity for LRS commissioning with the HV on was limited in this initial run.

### *3.5. Commissioning: Charge Readout*

During cryogenic commissioning but before the HV period, the charge readout was commissioned in multiple steps. First, a communication network between the ASICs of each pixel tile was established. At this stage, 43 out of 6400 ASICs were disabled due to unreliable communication or recurring bit-corruption issues. Next, a measurement of the baseline offset was performed by periodically sampling the charge on every pixel: this check quantifies the pedestal for each readout channel, and higher baseline noise in specific regions of the detector helps identify sources of noise impacting the charge readout. Once pedestals were quantified, the channel-level thresholds for the nominal self-triggering operations were tuned such that the trigger rate on each channel lay between 0.01 and 0.1 Hz. Finally, to estimate the threshold values, a special run was carried out where the periodic reset of the front end, typically configured with a period of 10–100 μs, was disabled such that the slowly integrating charge from leakage currents induced at-threshold triggers. The obtained thresholds were found to be 5 k electrons on average. Design and handling issues primarily related to system grounding, microphonics, and leakage currents ultimately required disabling 2.25% of all pixel channels over the course of these commissioning stages. The disabled channels were primarily located along the edge boundaries of the pixel tiles.

Design improvements implemented in a subsequent large-scale, ND-LAr prototype, the Full Scale Demonstrator (FSD), reduced the total fraction of inactive channels to 0.8%. High noise or leakage currents were responsible for a quarter of these disabled channels, while the remainder stemmed from chip communication issues.

### *3.6. Data Collection: Nominal HV with LAr Purity*

The 2x2 Demonstrator collected 86 h of NuMI beam data at nominal running conditions, with a field strength of 500 V/cm and a high level of LAr purity ($\mathcal{O}$(ms)). Additional data recorded at half-nominal field strength completes the physics data set, bringing the total run time with beam and high LAr purity to 4.5 days. This period of 100% detector up-time corresponded to $1.5 \times 10^{19}$ protons on target (POT), which produced more than 30k (anti)neutrino interactions in the LArTPC volume. A summary of the detector status and corresponding NuMI beam configuration is shown in Figure 9.

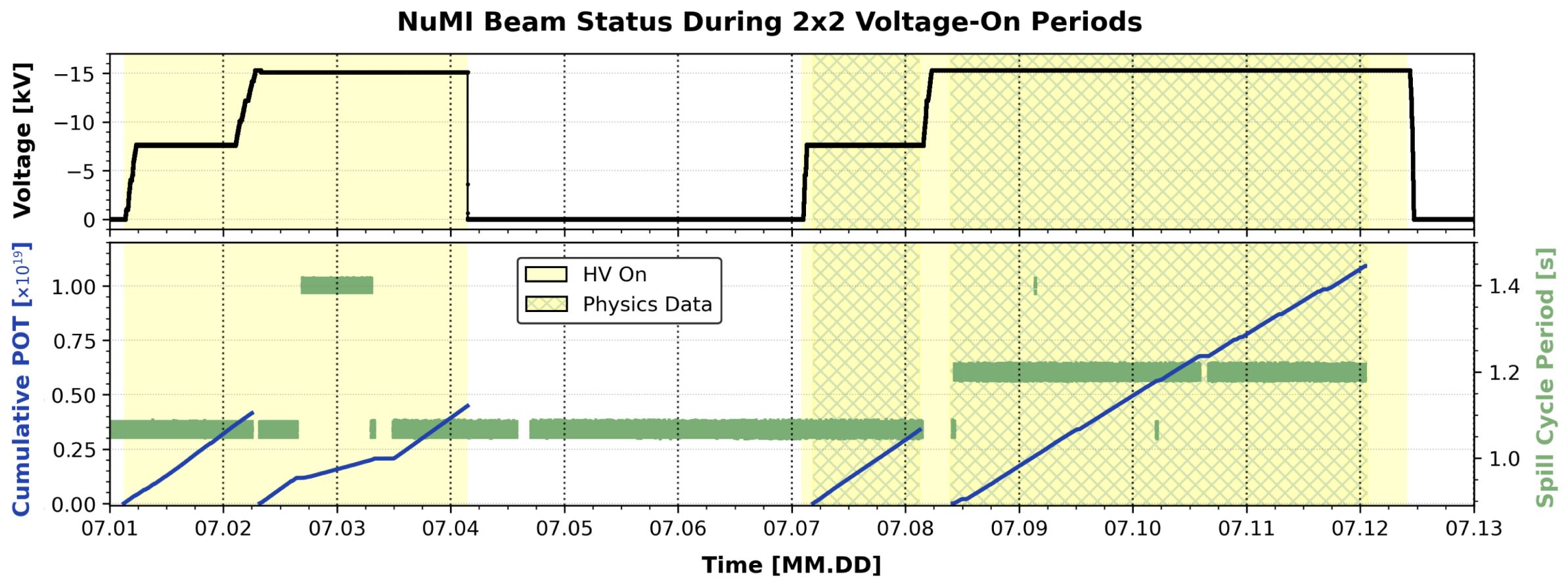

**Figure 9.** Recorded POT for the NuMI beam of the 2x2 Demonstrator (blue). The green line indicates when the beam was running and shows the cycle period of the beam pulses delivered. The topmost panel shows the TPC HV applied to the 2x2 cathode(s) in black.

Throughout the run, the Mx2 and LRS triggered on NuMI A9 early-warning signals to capture beam spills, while the LRS additionally triggered on off-beam interactions exceeding a set light threshold. The CRS pixels self-triggered continuously on independent, pixel-by-pixel thresholds; additional markers in the CRS data stream tagged NuMI A9 and LRS threshold triggers to assist in event building during downstream processing. Six hours of additional, low-threshold, non-beam data were collected for charge readout commissioning studies on 12 July 2024, after which the detector ramped down for further cryogenic improvements.

## 4. Data Validation and Event Displays

The innovative technology prototyped in the 2x2 Demonstrator—namely, its native 3D pixelated charge readout—coupled with a low-background, underground environment enables instantaneous proof of detector performance and the viability of the ND-LAr design. During the 4.5-day physics-quality run, LArTPC events could be viewed in event displays like those shown at the end of Section 4 almost immediately, using minimally processed raw data. The 2x2 Demonstrator consortium will publish data analyses utilizing offline event reconstruction at a later date. The remainder of this paper focuses on validations of trigger synchronization between detector subsystems and the NuMI beam, as well as event displays demonstrating the performance of the 2x2 Demonstrator with minimally processed data.

### 4.1. Validation: Multi-Detector Triggering

The three detector subsystems (LRS, CRS, and Mx2) record data through distinct DAQs in separate data streams. The event-level trigger alignment of each subsystem with the NuMI beam A9 early-warning signal, as well as each subsystem's timing offsets relative to each other, must be well-understood. Validation of the detector's trigger alignment is essential to ensure accurate offline data matching and event reconstruction.

Within the LArTPC, the charge readout and the light readout share a common on-beam triggering system. The light readout warm electronics receive the NuMI A9 directly and trigger, then forward the signal to a PACMAN controller to tag the spill window in the self-triggering charge readout. Mx2, in contrast, triggers independently, but also on the NuMI A9.

In order to optimize timing measurements within the 2x2 Demonstrator LArTPCs, the LRS has been configured to record light waveforms over a period of 16 μs. Each trigger is padded by approximately 1.6 μs prior to beam arrival, and the LRS continues taking data for several μs following the spill period to ensure full coverage of any associated secondary interactions or decays. The Mx2 likewise pads its beam window by a period of 0.5 μs and collects data for 5.5 μs following the end of the spill.

Figure 10 shows the integrated and normalized data rate spanning two hours from the three detector subsystems (CRS, LRS, and Mx2) relative to the arrival of the beam pulse. When compared to Figure 11, in which the 6-batch structure of the NuMI beam is visible, it is evident that all three subsystems successfully trigger on the beam. The beam structure is apparent in the shape of the Mx2 integrated waveforms and, to a lesser degree, in the light readout integrated waveforms.

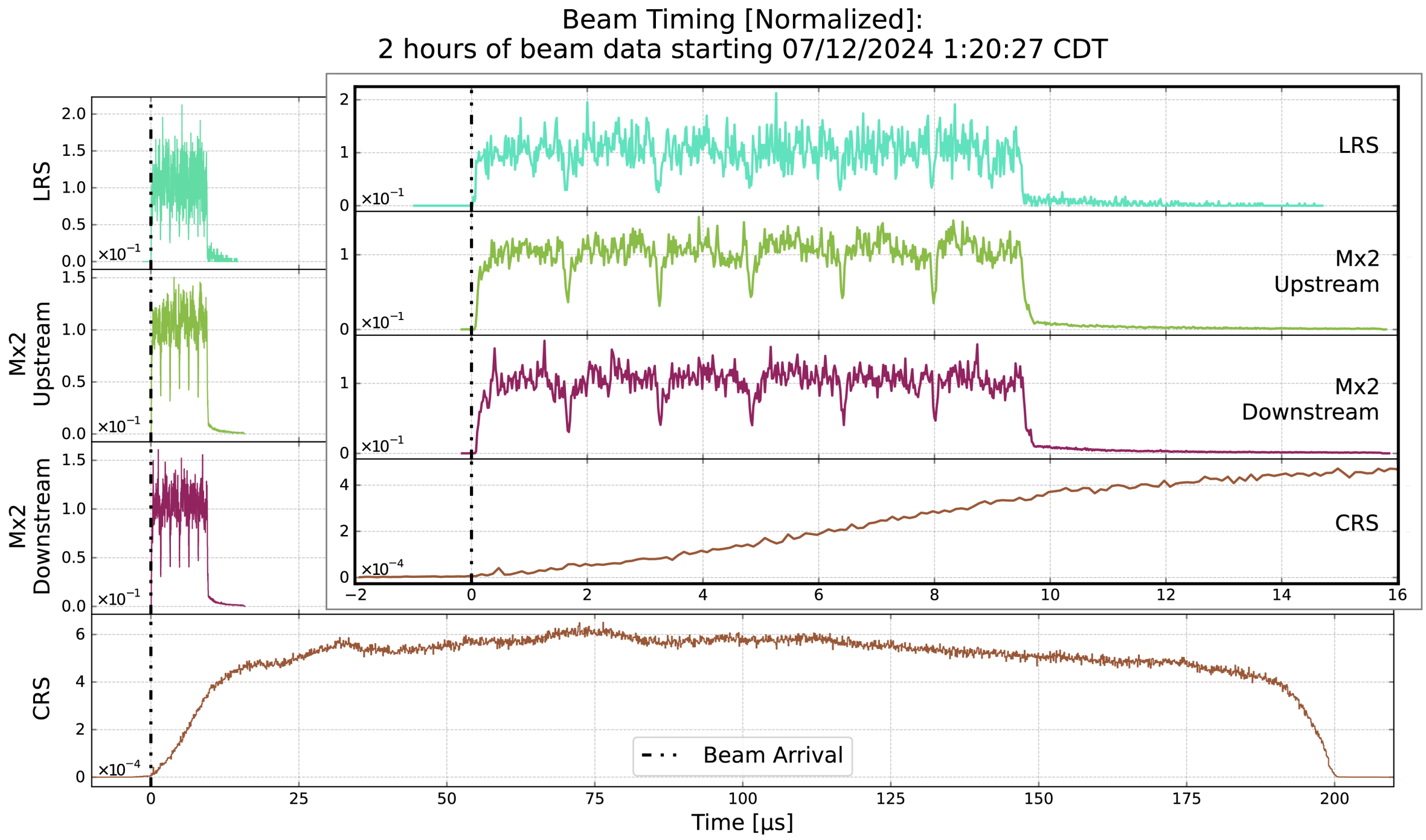

**Figure 10.** The maximum readout window of a single beam spill in the 2x2, 200 μs, is defined by the maximum drift time for charge. Above, the recorded signal outputs of the three subsystems of the 2x2 Demonstrator (LRS, Mx2, and the CRS), summed across two hours of beam spills, are plotted against the maximum readout window. Focusing on the first 16 μs (see inset), the successful alignment between subsystems is evidenced by the nearly simultaneous arrival time of the beam signal. Note that the LRS and Mx2 pad their readout by different lengths of time before and after the beam arrival, as determined by each systems' respective DAQ requirements.

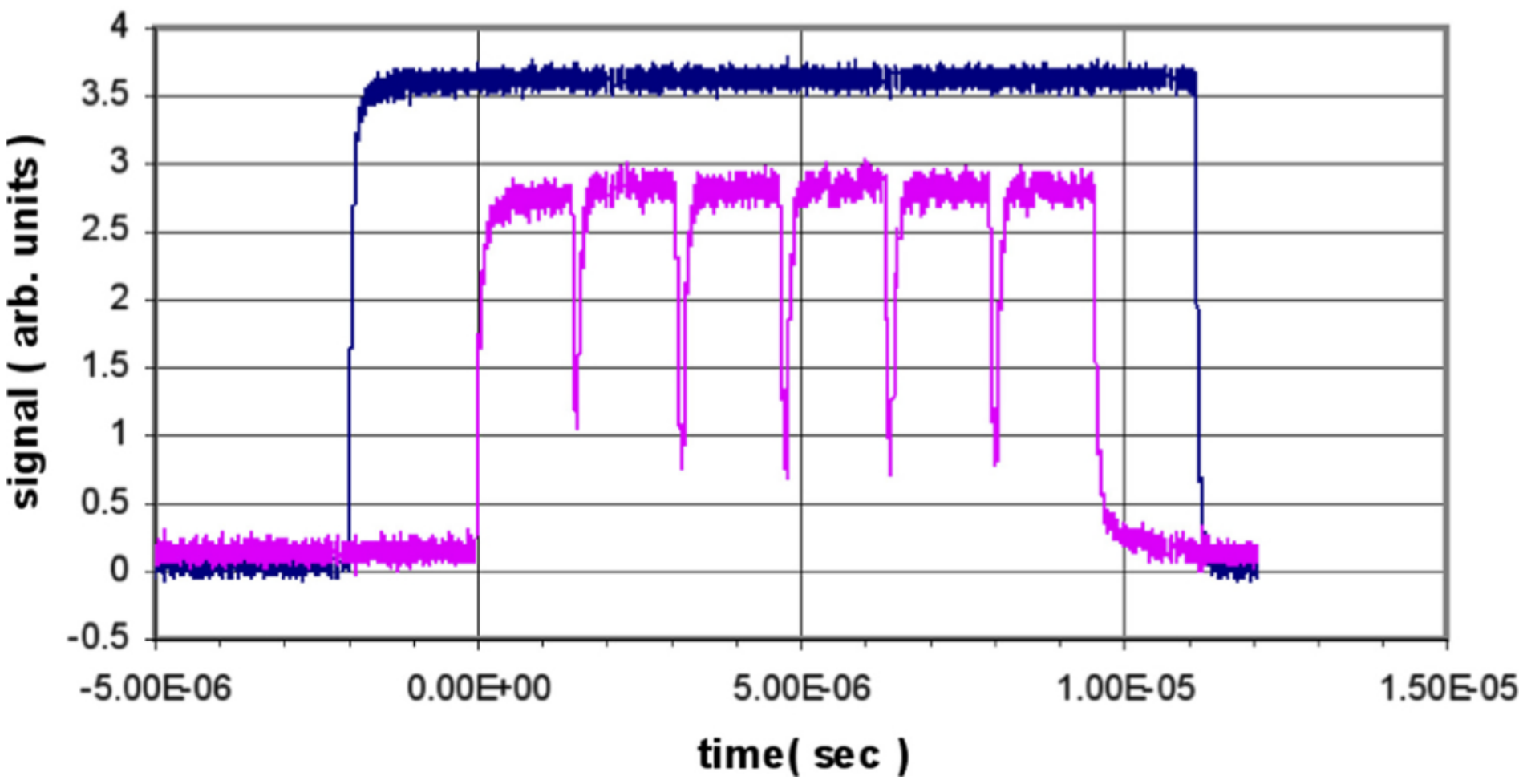

**Figure 11.** In magenta: the NuMI beam structure is evident in the NuMI toroid signal, with the beam running in slip-stacked, NuMI-only mode. In dark blue: NuMI's beam trigger window [6].

### 4.2. Validation: Charge Readout Self-Triggering

One advantageous feature of the 2x2 Demonstrator charge readout is its ability to self-trigger on events without reliance on an external beam or light threshold trigger. Outside of the ∼0.8 Hz NuMI beam spills, most high-energy interactions in the 2x2 Demonstrator stem from muons, often minimally ionizing, produced by cosmic ray interactions in the atmosphere. The MINOS ND, when located where the 2x2 Demonstrator currently stands, measured a cosmic muon rate of 27 Hz [44]. Although the MINOS target and the 2x2 target are materially different, the rock overburden is unchanged, so a scintillating detector with an approximate volume of 2 $m^3$ is estimated to detect cosmic muon events at a rate of around 2 Hz.

Figure 12 shows the self-trigger data rates for the charge readout system over several seconds, along with the $t_0$ of NuMI beam triggers. A clear correlation can be observed between self-triggered hits from the pixelated charge readout and the beam signal from NuMI. In addition, the non-beam-correlated spikes are consistent with the rate of cosmic muons expected in the 2x2 (∼2 Hz).

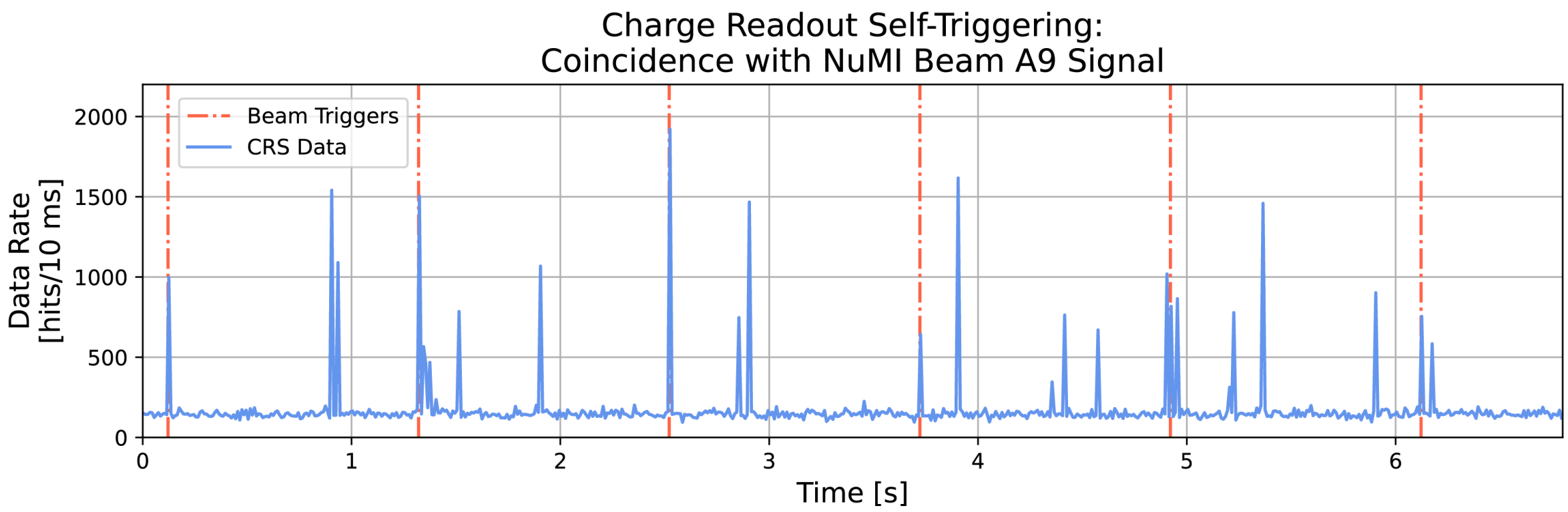

**Figure 12.** CRS self-trigger hit rates (blue) as a function of time, overlaid with A9 triggers from the NuMI beam (orange) indicating beam spill arrival times.

### 4.3. Visual Confirmation of Neutrino Interactions

With minimal offline processing, data recorded within a NuMI beam trigger can be visually analyzed to confirm the presence of neutrino interactions and the validity of the detector systems.

The 2x2 Demonstrator expects two consistent sources of relatively high-energy background events. The first, mentioned previously, are cosmic muons. These have no correlation to the beam timing. Due to the short duration of the NuMI beam pulse and the beam's comparatively long cycle time, beam events without any cosmic background are common. The second background source are muons produced by NuMI beam neutrino interactions in the rock between the beam target and the detector hall, called rock muons. These ionizing particles coincide with the beam spill. Simulations of rock muon interactions in the nearest 100 m of rock to the 2x2 Demonstrator suggest that an average of one to three rock muons will pass through the 2x2 LArTPC or its Mx2 muon-tagging planes during each beam pulse. Preliminary visual scanning of beam data supports this approximation, and a more detailed analysis is in progress.

The first events presented below, Events 1 and 2, were selected from our high-purity, beam-on dataset via visual scanning. These events feature data from all three subsystems. The data from the LArTPCs have not yet undergone full offline reconstruction; only minimal calibration has been applied to the raw data products. In each display, multiple charged current interactions occur in the LArTPC detector volume across one beam spill; these neutrino interactions are accompanied by multiple rock muon tracks visible in the external Mx2 planes, some of which enter the LArTPC. This density of events, while less than that expected at ND-LAr, provides an excellent test of the high-coverage LRS, the pixelated CRS, and the efficacy of detector modularity in separating simultaneous interactions.

Also included is an event display of a beam spill in which only rock muons deposit energy in the detector volume (Event 3), as well as a cosmic muon event recorded between beam spills (Event 4). These event displays provide insight into two prominent background topologies for both the 2x2 Demonstrator and ND-LAr.

The event displays presented in this paper feature a 3D projection of CRS and LRS data; an additional 3D projection includes particle trajectories passing through the upstream and downstream Mx2 planes when appropriate. Light waveforms have been corrected for baseline offsets, and SiPM responses have been equalized and converted from ADC counts to detected photoelectrons using the calibrations reported in Section 3. The following event displays use calibrated light data to render ACL and LCM responses visually coherent. Given the impracticality of displaying full light waveforms from all 384 SiPMs, the calibrated sum of each SiPM waveform is represented in the 3D projections as a colored panel. The colored panels extend along the drift axis to cumulatively fill the geometric area occupied by TPB-coated, ACL or LCM light traps within the detector. Although multiple SiPMs are coupled to each light trap, photon transport is not uniform within the traps themselves and the signal observed by each coupled SiPM depends upon its proximity to the source of scintillation light. Each SiPM sum is displayed independently to better demonstrate the systems' resultant capability for spatial discrimination.

Two-dimensional projections of both the charge and light signals for each event are also provided in each display. Three projections—one from above the detector, one side-on along the beam axis, and the last side-on along the drift axis—show the charge and light signal summed along the third, flattened axis. Light signal sums are represented by colored rectangles running along the borders of each TPC. These rectangles do not extend along the drift axis in the side-on projection in order to avoid obscuring the charge data.

**Event 1: NuMI beam trigger on contained charged current muon neutrino.**
This event shows a charged current neutrino interaction with its vertex in Module 3, upstream. The interaction produces one through-going muon track which passes above the downstream Mx2 planes, as well as several other tracks that are contained within the LArTPC volume.

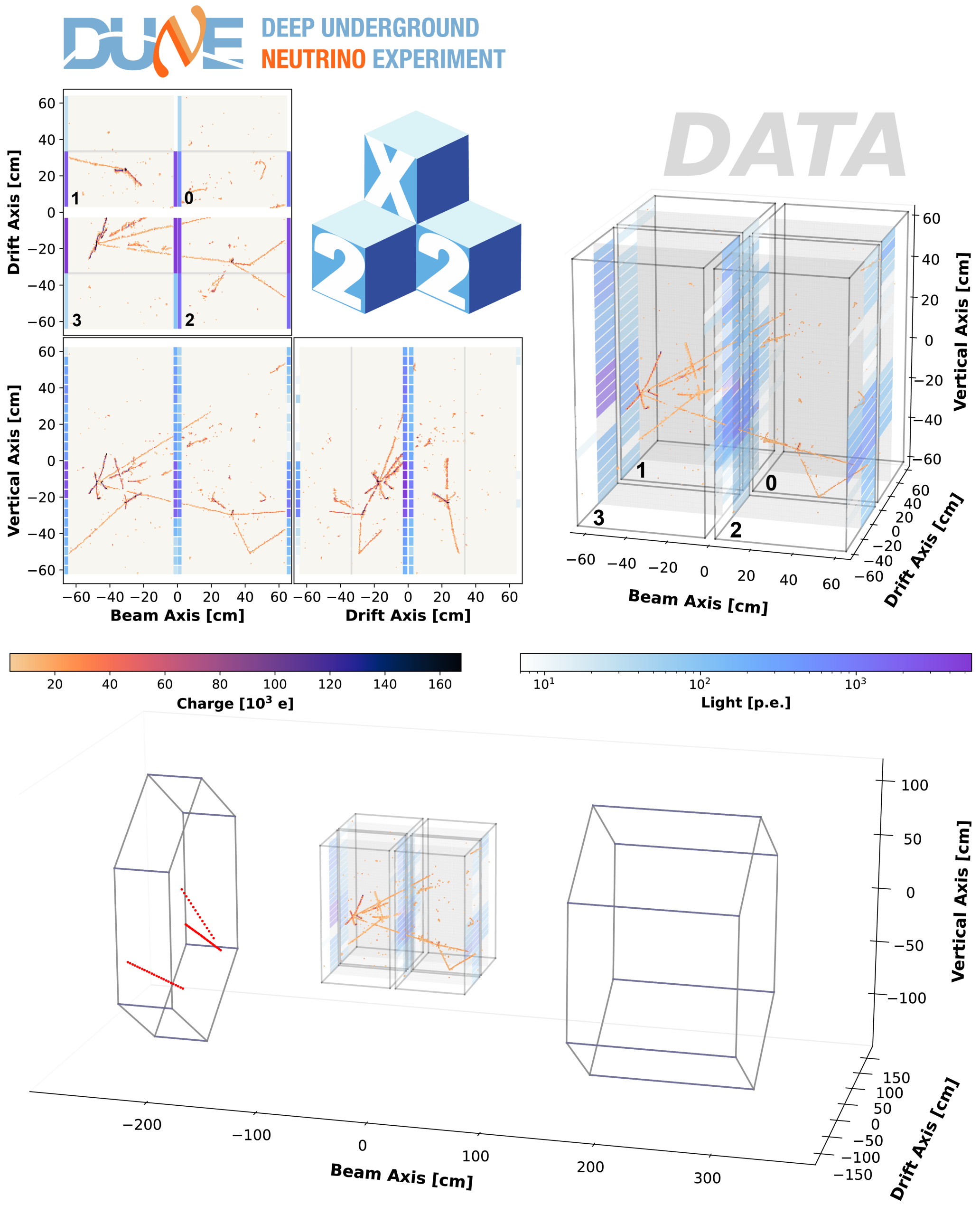

**Event 2: NuMI beam trigger on two charged current neutrino interactions.**
Two charged current neutrino interactions, one in Module 1 and the other in Module 2, produce tracks within the LArTPC volume. A rock muon passing through both upstream and downstream Mx2 taggers pierces Modules 2 and 3 during the same period of 16 μs, but the neutrino interactions remain well separated from each other by the detector's modularity.

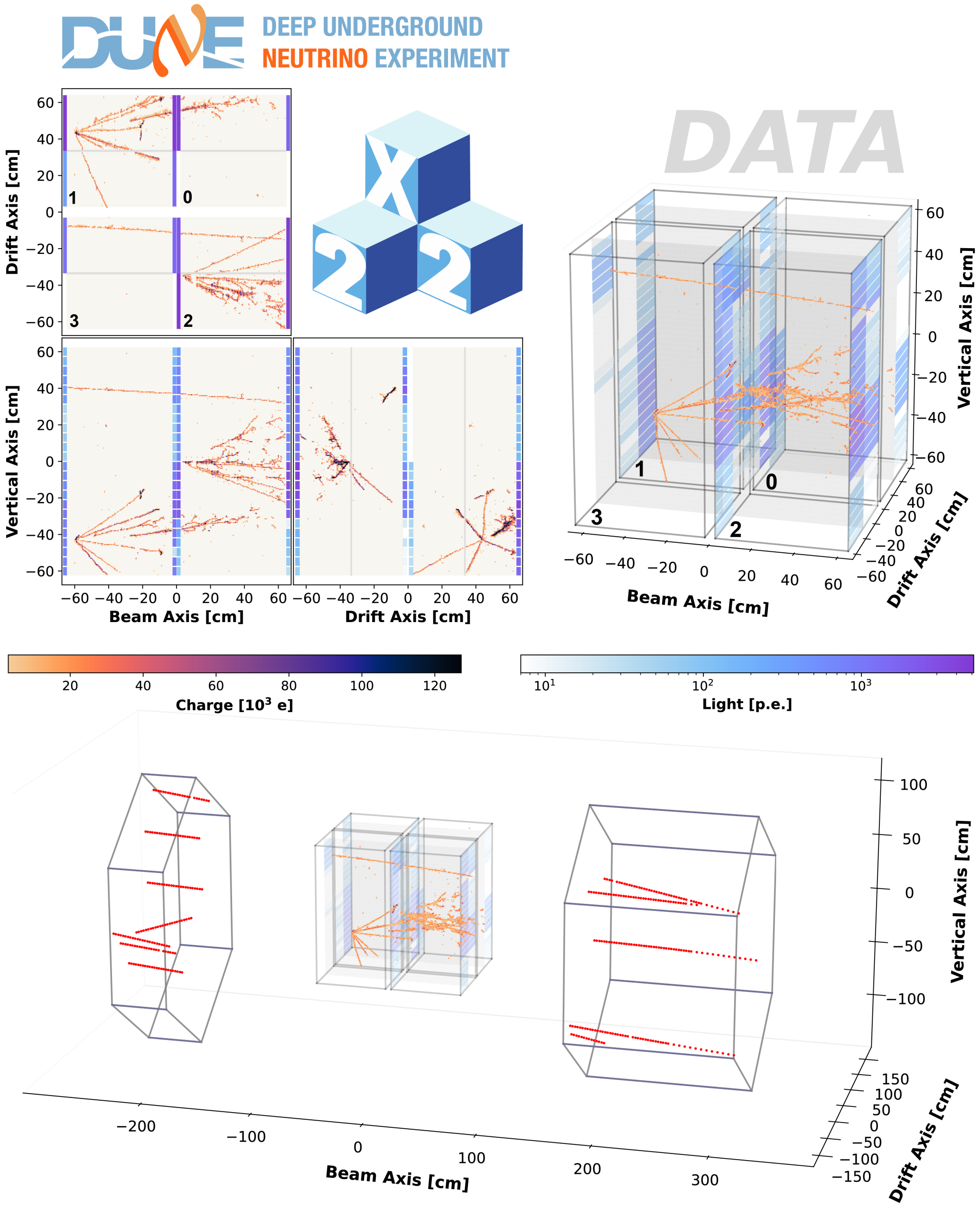

**Event 3: NuMI beam trigger on external neutrino interactions with Michel electrons.** Several rock muons produce tracks in the LArTPC volume; one of these muons decays in Module 3, generating a Michel electron. In the light waveforms corresponding to the LCMs nearest the decay (top left), one can see a separation of 2.48 µs between fast scintillation signals associated with the muon and the Michel electron.

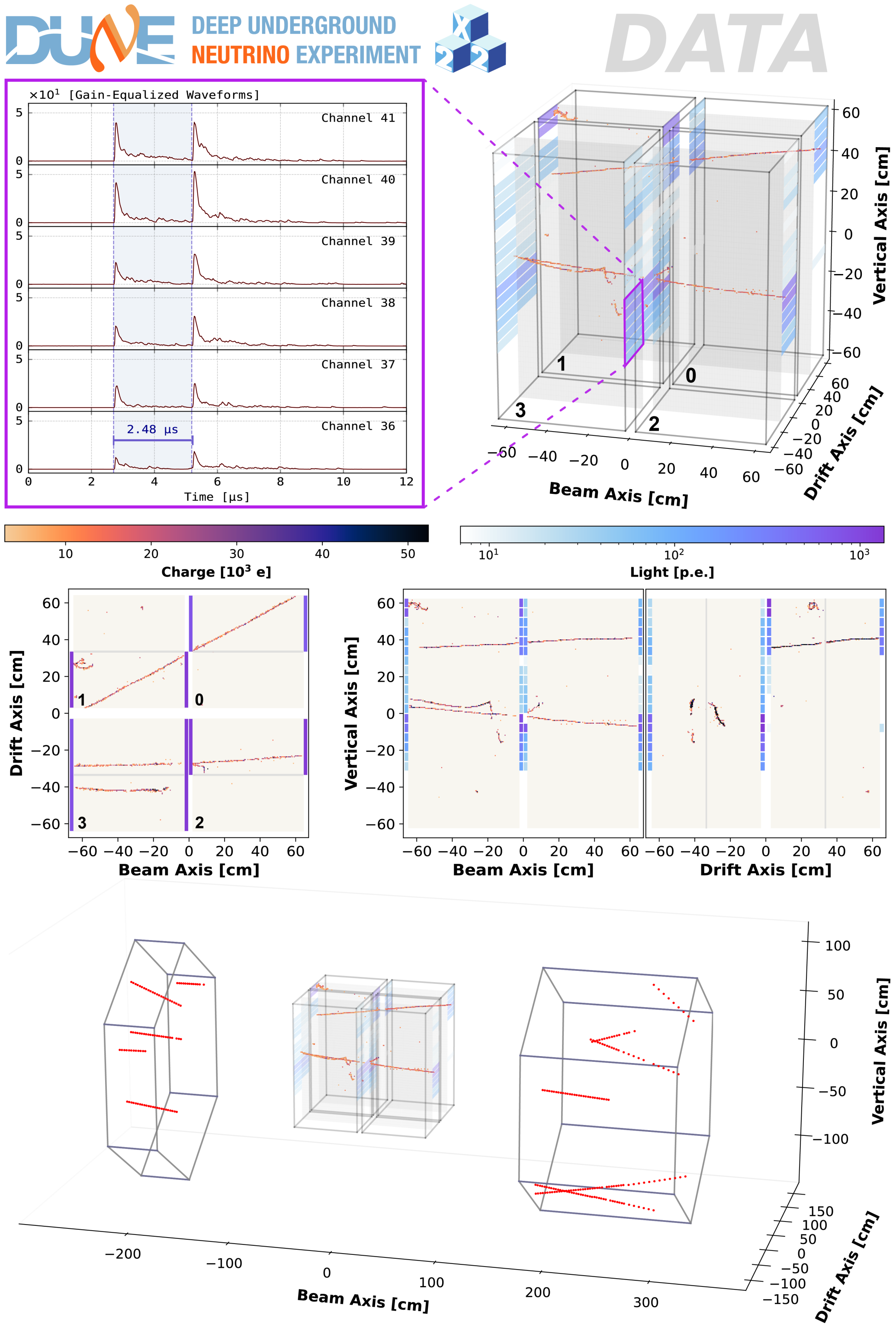

**Event 4: Light threshold trigger on cosmic muon.**
A cosmic muon enters the LArTPC at a relatively shallow angle; with only the minimally-processed pixel hits, showering tracks are clearly defined, while the y-axis spatial resolution of the LRS is visible in the waveform sums bordering each charge display. Additionally, as the cosmic muon only deposits energy in Modules 2 and 3, the successful optical isolation of adjacent TPCs is clearly demonstrated. As the external Mx2 panels trigger solely on the NuMI A9 early-warning signal, there are no recorded Mx2 tracks corresponding to this off-beam LArTPC event.

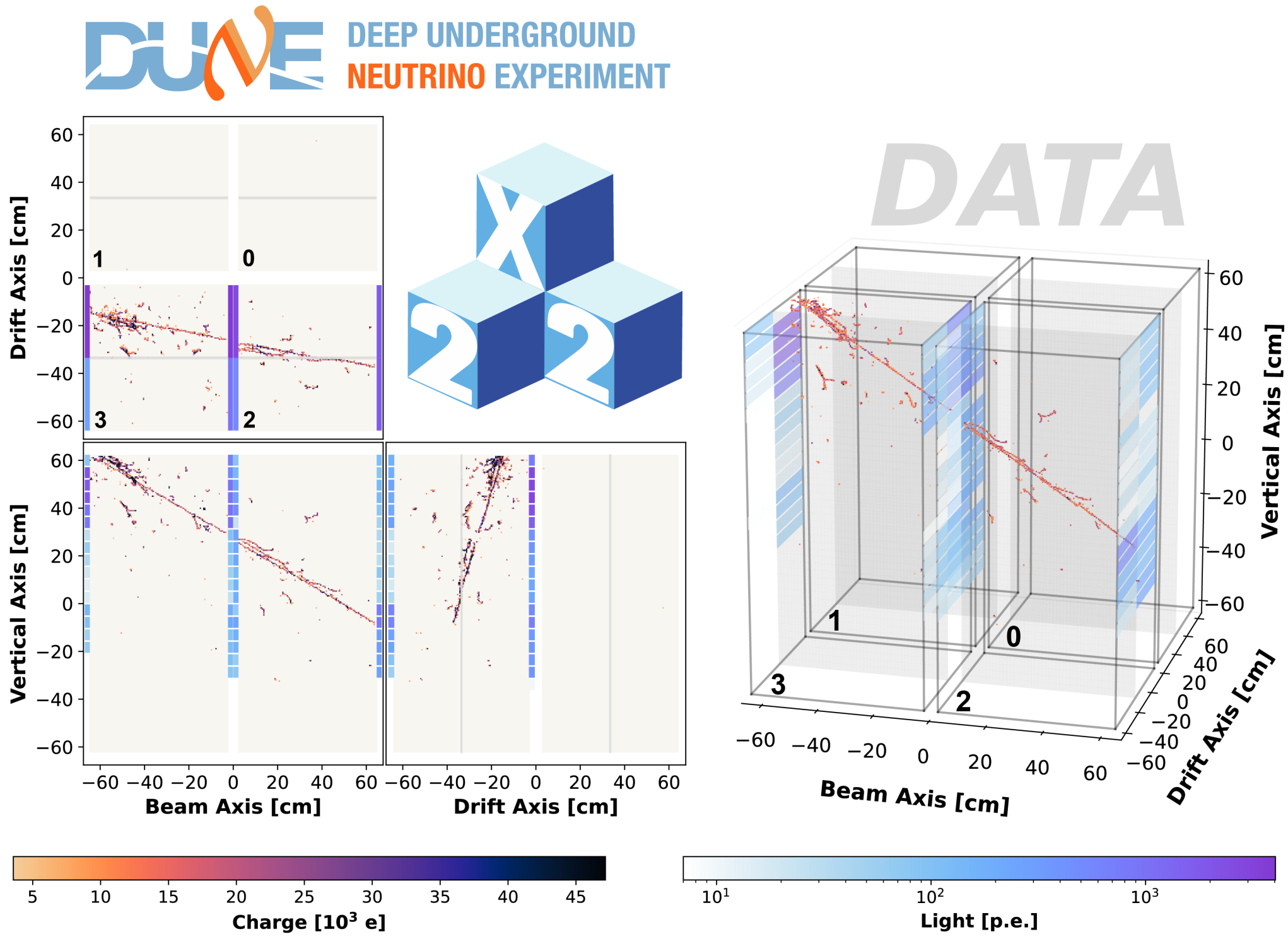

## 5. Lessons Learned

The 2x2 Demonstrator, as an intermediate stage in the ND-LAr prototyping process, has informed updates to the intended ND-LAr design. Several of these updates have already been tested in the FSD, a single ND-LAr-scale module assembled and run at the University of Bern. The FSD was successfully operated at a nominal drift field of 500 V/cm during October and November 2024. The FSD prototyped shielding on the LRS cold electronics, which improved the isolation between the charge and light subsystems. It also featured updated warm and cold cabling for the LRS, as well as a new iteration of the pixelated anode tiles. These alterations reduced the noise levels observed in the LArTPC data, particularly for the LRS. Although the resistive field shell described in Section 2 was, and continues to be, successfully prototyped in the 2x2, cost constraints motivated its replacement by an equally low-profile resistor-chain field shell in subsequent prototypes. The resistor-chain design was successfully operated in 2024 as part of the FSD, and its performance satisfies ND-LAr physics requirements. Future testing in the 2x2 Demonstrator and the FSD will aim to streamline calibration methods, further reduce noise levels, and improve the dynamic range of the LRS. Additionally, the Module Row Prototype, still under construction at Fermilab, will test the mounting, structural support, and installation of five full-scale modules ahead of ND-LAr assembly.

## 6. Conclusions

Between 11 June and 12 July 2024, the 2x2 Demonstrator of the DUNE Near Detector LArTPC was commissioned and collected 10 days of on-axis NuMI beam data at Fermilab. Of the acquired data, 86 h of the NuMI triggered beam data were collected at nominal running conditions, with a field strength of 500 V/cm and an LAr purity of approximately 1.25 ms. In total, the 2x2 Demonstrator collected around 30,000 neutrino interactions during its physics-quality run. Although the data collected by the 2x2 Demonstrator will be processed through customized reconstruction [45,46] and analysis tools, the clarity of event topologies in the July 2024 dataset enable us to identify and display the first recorded neutrino events in a DUNE prototype detector using minimally processed subsystem signal data.

The 2x2 Demonstrator is expected to resume collecting neutrino data when the NuMI beam returns. In the interim, the existing dataset is being utilized as a training tool for upcoming analyses. In addition, these data will help to advance the development of native 3D reconstruction algorithms, to validate the proposed design of the DUNE Near Detector, and to inform the development of future LArTPC detectors.

**Funding:** This document was prepared by the DUNE collaboration using the resources of the Fermi National Accelerator Laboratory (Fermilab), a U.S. Department of Energy, Office of Science, HEP User Facility. Fermilab is managed by Fermi Forward Discovery Group, LLC, acting under Contract No. 89243024CSC000002. This work was supported by CNPq, FAPERJ, FAPEG and FAPESP, Brazil; CFI, IPP, and NSERC, Canada; CERN; MŠMT, Czech Republic; ERDF, H2020-EU, and MSCA, European Union; CNRS/IN2P3 and CEA, France; INFN, Italy; FCT, Portugal; NRF, Republic of Korea; CAM, Fundación "La Caixa", Junta de Andalucía-FEDER, MICINN, and Xunta de Galicia, Spain; SERI and SNSF, Switzerland; TÜBİTAK, Turkey; The Royal Society and UKRI/STFC, United Kingdom; DOE and NSF, United States of America. This research used resources of the National Energy Research Scientific Computing Center (NERSC), a U.S. Department of Energy Office of Science User Facility operated under Contract No. DE-AC02-05CH11231. This manuscript has been authored by Fermi Forward Discovery Group, LLC, under Contract No. 89243024CSC000002 with the U.S. Department of Energy, Office of Science, Office of High Energy Physics.

**Data Availability Statement:** The datasets presented in this article are not readily available because the data are subject to restrictions per the DUNE collaboration policies and Data Management Plan. Inquiries regarding the data may be directed to the corresponding author.

**Conflicts of Interest:** The authors declare no conflicts of interest.